\documentclass{llncs}
\usepackage{makeidx}  
\usepackage[square,comma,numbers,sort&compress]{natbib}
\usepackage{graphicx}
\usepackage[dvips]{color}       
\usepackage{bm}                 
\usepackage{amsmath}            
\usepackage{amsfonts}
\usepackage{amssymb}
\usepackage{bbm}
\usepackage{latexsym}           

\begin{document}
\frontmatter          
\pagestyle{headings}  

\mainmatter              
%
\title{Spatial search using the discrete time quantum walk}
%
\titlerunning{Spatial search with quantum walks}  
%
\author{Neil B. Lovett \and Matthew Everitt \and Matthew Trevers \and Daniel Mosby \and Dan Stockton \and Viv Kendon}

\authorrunning{Neil Lovett et al.} 

\institute{School of Physics and Astronomy, University of Leeds, Leeds, LS2 9JT,
 UK\\
\email{pynbl@leeds.ac.uk, V.Kendon@leeds.ac.uk}\\
}

%
\maketitle              

\begin{abstract}
We study the quantum walk search algorithm of \citeauthor*{shenvi02a}
[PRA 67 052307 (2003)] on data structures of one to two spatial
dimensions, on which the algorithm is thought to be less efficient than in three
or more spatial dimensions.  Our aim is to understand
why the quantum algorithm is dimension dependent whereas the best
classical algorithm is not, and to show in more detail how the
efficiency of the quantum algorithm varies with spatial dimension
or accessibility of the data. Our numerical results agree with the expected scaling in 2D of $O(\sqrt{N \log N})$, and show how the prefactors display significant dependence on both the degree and symmetry of the graph. Specifically, we see, as expected, the prefactor of the time complexity dropping as the degree (connectivity) of the structure is increased.
\end{abstract}

\section{Introduction}
\label{sec:intro}

The promise of quantum computers to provide fundamentally faster
computation is dependent both on being able to build a quantum
computer of a useful size, and on finding algorithms it can run
that are faster than any classical algorithm.
Since some of the most efficient classical algorithms are based
on random walks, a natural place to look for faster quantum 
algorithms is to see if there is a faster quantum version of
a random walk.  This approach was pioneered by 
\citeauthor{ambainis01a}~\cite{ambainis01a} and
\citeauthor{aharonov00a}~\cite{aharonov00a}, who proved
that a discrete time quantum walk on the line or cycle spread or mixed, 
respectively, quadratically faster than a classical random walk.
Continuous time quantum walks were introduced by 
\citeauthor{farhi98a}~\cite{farhi98a} and provide the same
algorithmic speed up as discrete time quantum walks.  We concentrate
on the discrete time walk in this paper, since it is more convenient
for numerical calculations.

One of the first algorithmic applications of quantum walks was 
searching of unsorted databases.  \citeauthor{shenvi02a}~\cite{shenvi02a}
proved a discrete quantum walk can replicate the performance of Grover's
search algorithm \cite{grover96a} by finding a marked item among
a total of $N$ in $O(\sqrt{N})$ steps.
The quantum walk search is now a standard tool in the development
of quantum algorithms, for reviews
see \citeauthor{ambainis04a}~\cite{ambainis04a} and
\citeauthor{santha08a} \cite{santha08a}.

In this work, we are interested in the case where the data to be
searched is also physically restricted in how it can be accessed.
For example, to read a particular item from a magnetic tape, it is
necessary to wind through the tape until the correct position is 
reached.  Data stored on a hard disk is arranged in concentric rings
on spinning disks: the heads move sideways across the rings and then
the item can be read within the next revolution of the disk.
This is quicker than reading data on a magnetic tape, but still
requires a significant amount of time per item.
A classical search of data stored on tape
has to work through the tape from one end to the other, testing each
item in turn to see if it is the required one.
In fact, this is the optimal classical strategy for any
arrangement of unsorted data on any storage medium.
Finding a particular item requires in the worst case all the
items to be checked, and on average half of the items will be
examined before the marked one is located.  This gives a classical
running time of $O(N)$.

The quantum walk search algorithm of \citeauthor{shenvi02a}~\cite{shenvi02a}
arranged the data to be searched as the nodes of a graph on which 
the quantum walk then propagated.  Specifically, they used a
hypercube (of dimension $\lceil\log_2 N\rceil$), for which the
quantum walk can be solved analytically \cite{moore01a}.
They proved their quantum walk search can equal the quadratic speed up over
classical searching given by Grover's search algorithm \cite{grover96a}
-- this is known to be optimal \cite{bennett97a}.
Improvements by \citeauthor{potocek08a}~\cite{potocek08a} bring the
actual running time of the quantum walk search algorithm
very close to the optimal one.

The hypercube is a highly connected structure, which doesn't correspond
to physically realistic storage media.  Motivated by this observation,
study of lower dimensional search began with
\citeauthor{benioff00a}~\cite{benioff00a}, who considered the additional
cost of the time taken for a robot searcher moving between different
spatially separated data items.
\citeauthor{aaronson03a}~\cite{aaronson03a} then produced a
quantum algorithm that finds a marked item in $O(\sqrt{N})$ for
data arranged on lattices of dimension greater than two, and
$O(\sqrt{N} \log^{3/2}N)$ for a square lattice (dimension $D=2$).
Work by \citeauthor{childs03a}~\cite{childs03a,childs04a} and
\citeauthor{ambainis04b}~\cite{ambainis04b}
found quantum walk algorithms with the same running time
for $D>2$ and a small improvement to $O(\sqrt{N}\log N)$ for $D=2$.
\citeauthor{tulsi08a} \cite{tulsi08a} recently improved this again
to $O(\sqrt{N\log N})$ with a modified approach using ancilla qubits.
Recent results from \citeauthor{magniez08a} \cite{magniez08a} have shown that this result is unlikely to be improved. They show that the hitting time of a quantum walk is quadratically faster than the hitting time of a classical random walk for classical walks which are reversible. They prove this speed up is actually tight and cannot be improved upon for a large class of quantum walks where the unitary is a reflection. The classical hitting time on a 2D lattice using a reversible random walk is $O(N \log N)$. We can see how the $O(\sqrt{N \log N})$ run time of \cite{tulsi08a} fits this result exactly. In his recent work, \citeauthor{tulsi08a} is able to find the marked state with constant probability, $O(1)$, using a modified version of the \citeauthor{shenvi02a} search algorithm with ancillas. \citeauthor{magniez08a} \cite{magniez08a} extend this work and show how to find the marked state with constant probability in the same improved time, $O(\sqrt{N \log N})$, for any quantum walk based on a reversible, ergodic (a stationary distribution can be found) classical random walk. These results and others in recent papers from \citeauthor{santha08a} \cite{santha08a} and \citeauthor{patel10a} \cite{patel10a} show it is unlikely that this extra $\sqrt{\log N}$ factor in the run time can be removed. 

When we first started this work, our aim was to investigate whether the lower bound of $O(\sqrt{N} \log N)$ previously shown for the 2D lattice was tight. We were looking to gain insight into what happens between $D=1$, where simple arguments show that no speed up is expected, and $D=3$, where
the quantum speed up is known to be optimal. Since then, \citeauthor{magniez08a} \cite{magniez08a} and \citeauthor{krovi10a} \cite{krovi10a} have proved that it wasn't and have shown the optimal lower bound on a 2D lattice is $O(\sqrt{N \log N})$, which our results confirm. We were also interested in how dependent the prefactors of the runtime were on the connectivity of the structure. Using numerical techniques, we have been able to study other regular two dimensional structures with varying connectivity. Our results on these structures also confirm the two dimensional scaling of $O(\sqrt{N \log N})$, but show how the prefactors change in relation to connectivity.

The quantum walk search algorithm has a strong dependence on the spatial dimension of the structure to be searched. The best known classical search algorithms show no such dimensional dependence. These classical algorithms are not based on classical walks but just search each item in sequence, until the marked state is found. Search algorithms based on classical walks are not optimal, for example, a random walk on a cycle would have a run time of $O(N^2)$. Quantum walk search algorithms have been proved by \citeauthor{magniez08a} \cite{magniez08a} to be quadratically faster than classical searches by random walks which are reversible and ergodic. \citeauthor{krovi10a} \cite{krovi10a} extended this proof to require only that the 
classical random walk be reversible. In the example of the walk on the cycle, reversibility means the walk has a probability to propagate in both clockwise and anti-clockwise directions. The dimensional dependence of quantum walk searching thus follows a similar dependence in 
classical random walk based search algorithms. 

We are also interested in how important the symmetry of the database arrangement is for searching using quantum walks.  For a quantum walk crossing a hypercube, or the special ``glued trees'' graph of \cite{childs02a}, it is known that defects in the graph can seriously disrupt the efficiency of the quantum walk \cite{tregenna03a,krovi05a,krovi06a,krovi07a}. The marked state is treated as a type of symmetry breaking in the dynamics of the quantum walk, so other distinguished parts of the data structure (such as the end of the tape, or the edge of the hard disk) might distract the quantum walk from finding the desired marked state.

We begin in section \ref{sec:line} by reviewing the quantum walk on the 
line as introduced by \citeauthor{ambainis01a} \cite{ambainis01a}.  We describe how 
the coin toss operation can be varied to provide the broken 
symmetry for the marked state, as was done by \citeauthor{shenvi02a}~\cite{shenvi02a} in their original quantum walk search algorithm.  We 
summarise the basic properties of the quantum walk on the 
line, to introduce the concept of quantum walks and how they 
differ from classical random walks.  We then move on in 
section \ref{sec:2Dsearch} to quantum walks on a two-dimensional Cartesian 
lattice, in order to review the quantum walk search 
algorithm of \citeauthor{shenvi02a}~\cite{shenvi02a} in a context where it gives a 
good speed up over the best classical algorithms.  This has 
been solved analytically by \citeauthor{ambainis04b} \cite{ambainis04b}. By using 
numerical simulation, this allows us to explore the actual 
scaling with the size of the system and compare this with 
the bounds proved previously \cite{aaronson03a,ambainis04b}. We also varied the 
form of the coin operator applied to the marked state, to 
explore the sensitivity of the algorithm to the exact coin 
operator used.

We return to the line in section \ref{sec:1Dsearch} to explore the behaviour 
of quantum walk search algorithms in one spatial dimension. 
Previous work by \citeauthor{szegedy04a} \cite{szegedy04a} using his Markov chain based 
quantum walk shows that it only `finds' the marked state in 
time $O(N)$ with probability $1/N$. Obviously, this is of no use as every other state on the line will also have probability $1/N$, just like the original superposition. We confirm this numerically 
for the \citeauthor{shenvi02a}~\cite{shenvi02a} quantum walk search algorithm, 
describing in detail how boundaries and different forms of 
the coin operators for both the marked state and general 
evolution of the quantum walk affect the behaviour.  In 
section \ref{sec:2dstructures} we apply our numerical studies to structures with 
a different degree at each vertex, while still remaining in 
two spatial dimensions, namely the hexagonal lattice (each 
vertex having degree 3), a 2D Cartesian lattice with 
diagonals added (degree 8), and a Bethe lattice of degree 3. 
We report in detail how the quantum walk search algorithm 
success probabilities and run times vary compared with the 
plain 2D Cartesian lattice studied in section \ref{sec:2Dsearch}.  Finally, 
we discuss our results and plans for future work in section 
\ref{sec:conc}.

\section{Quantum walk on the line}
\label{sec:line}

\begin{figure}[!tb]
\begin{minipage}{\columnwidth}
    \centering
	\includegraphics[scale=0.4]{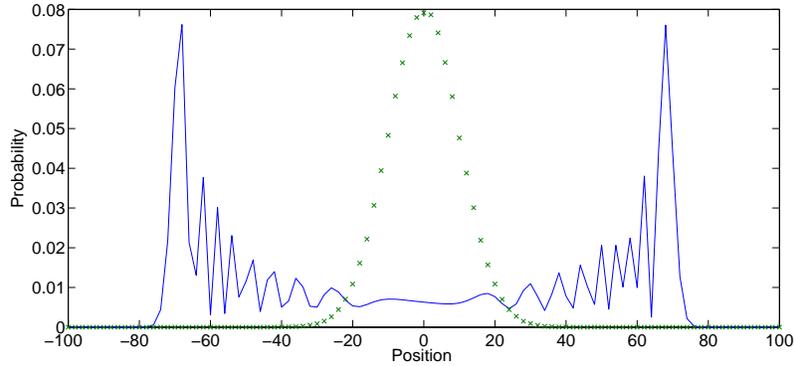}
	\caption{Classical (crosses) and quantum (solid line) probability
	distributions for walks on a line after 100 timesteps.
	Only even positions are shown since odd positions are zero.
	Initial state for the quantum walk of $(|0\rangle+i|1\rangle)/\sqrt{2}$.}
    \label{qandcwalk}
\end{minipage}
\end{figure}

A discrete time quantum walk on the line is defined in direct analogy with
a classical random walk: there is a walker carrying a coin
which is tossed each time step and the walker steps left or
right according to the heads or tails outcome of the coin toss.
We denote the basis states for the quantum walk as an ordered pair of
labels in a ``ket'' $|x,c\rangle$, where $x$ is the position and
$c\in \{0,1\}$ is the state of the coin.
A unitary coin operator is used at each time step and then a shift operation
is applied to move the walker to its new positions.
The simplest coin toss is the Hadamard operator $H$, defined by its action
on the basis states as
\begin{align}
H \mid x, 0 \rangle &= \frac{1}{\sqrt{2}}(\mid x, 0 \rangle + \mid x, 1 \rangle) \nonumber \\
H \mid x, 1 \rangle &= \frac{1}{\sqrt{2}}(\mid x, 0 \rangle - \mid x, 1 \rangle),
\label{hadamard}
\end{align}
and the shift operation $S$ acts on the basis states thus
\begin{align}
S \mid x, 0 \rangle &= \mid x-1, 0 \rangle \nonumber \\
S \mid x, 1 \rangle &= \mid x+1, 1 \rangle.
\label{shift}
\end{align}
The first three steps of a quantum walk starting from the origin are as follows
\begin{align}
(SH)^3\mid 0, 0 \rangle &=  (SH)^2 S \frac{1}{\sqrt{2}}(\mid 0, 0 \rangle \ + \mid 0, 1 \rangle) \nonumber \\
&= (SH)^2 \frac{1}{\sqrt{2}}(\mid -1, 0 \rangle \ + \mid 1, 1 \rangle) \nonumber \\
&= (SH) S \frac{1}{2}(\mid -1, 0 \rangle \ + \mid -1, 1 \rangle \ + \mid 1, 0 \rangle \ - \mid 1, 1 \rangle) \nonumber \\
&=  SH \frac{1}{2}(\mid -2, 0 \rangle \ + \mid 0, 1 \rangle \ + \mid 0, 0 \rangle \ - \mid 2, 1 \rangle) \nonumber \\
&= S \frac{1}{\sqrt{8}}(\mid -2, 0 \rangle \ + \mid -2, 1 \rangle \ + \mid 0, 0 \rangle \ - \mid 0, 1 \rangle \ + \mid 0, 0 \rangle \ + \mid 0, 1 \rangle \ - \mid 2, 0 \rangle \ + \mid 2, 1 \rangle) \nonumber \\
&= \frac{1}{\sqrt{8}}(\mid -3, 0 \rangle \ + \mid -1, 1 \rangle \ + 2 \mid -1, 0 \rangle \ - \mid 1, 0 \rangle \ + \mid 3, 1 \rangle).
\label{3steps}
\end{align}
As the walk progresses, quantum interference occurs whenever there is more
than one possible path of $t$ steps to the position.
This can be both constructive and destructive, as shown in eq.~(\ref{3steps}),
which causes some probabilities to be amplified or decreased at
each timestep.
This leads to the different behaviour compared to its classical counterpart:
the position of a walker following a classical random walk on a line 
spreads out in a binomial distribution about its starting point.
Both classical and quantum distributions are illustrated after 100 steps
in fig.~\ref{qandcwalk}.
It is clear in fig.~\ref{qandcwalk} that the quantum walk spreads faster.
\citeauthor{ambainis01a}~\cite{ambainis01a} proved that the quantum walk
spreads in $O(t)$ compared to a classical random walk which spreads
in $O(\sqrt{t})$.

Adding some decoherence into this basic walk \cite{kendon06a}
actually helps for some applications.
A nearly smooth ``top hat'' distribution can be obtained with a small amount of
decoherence, which is useful for random sampling in Monte-Carlo simulations
as an example. 
This also gives us a clue about how to use a quantum walk for searching.
The ``top-hat'' is obtained after starting at a single location.
Pure unitary dynamics are reversible, so
if we run the quantum walk backwards from a uniform distribution
on all points, we might expect it will go
to being approximately located at a single point.
Of course, it doesn't know which single point to converge on unless
we mark it in some way: we do this with a different coin operator.
If the coin operator allows the marked state to retain more
probability than it passes on, this creates a bias in the walk at this point.
This can be done with a biased Hadamard coin operator
\begin{equation}
H_{\delta} = \left( \begin{matrix} \sqrt{\delta} & \sqrt{1-\delta} \\ \sqrt{1-\delta} & -\sqrt{\delta} \end{matrix} \right ),
\label{biasedcoin}
\end{equation}
where $\sqrt{.}$ is the positive root,
and $H_{\delta}$ acts only on the coin state.
The value of $\delta$ determines how much of the incoming probability
is sent in each direction.  Taking $\delta=1$ gives the $\sigma_{z}$ operation;
the unbiased Hadamard operator eq.~(\ref{hadamard})
corresponds to $\delta = 1/2$; and
$\delta=0$ gives the $\sigma_x$ (spin flip) operator,
\begin{equation}
\sigma_{x} = \left( \begin{matrix} 0 & 1 \\ 1 & 0 \end{matrix} \right ).
\label{reflectcoin}
\end{equation}

Since this search algorithm doesn't provide any speed up on the line, we
next describe how it works on a two-dimensional Cartesian lattice.
The behaviour of the quantum walk search on a line will
then be compared in section \ref{sec:1Dsearch}.

\section{Two-dimensional Cartesian lattice}
\label{sec:2Dsearch}

A quantum walk in a higher dimension needs a larger coin, 
with one coin dimension for each choice of direction at the 
vertices.  The shift operator is enlarged in a similar 
straightforward way from the version in eq.~(\ref{shift}). As in 
the quantum walk search algorithm of \citeauthor{shenvi02a} \cite{shenvi02a}, as 
solved by \citeauthor{ambainis04b} \cite{ambainis04b} for the 2D Cartesian lattice, 
we use a symmetric coin operator based on Grover's diffusion 
operator,
\begin{equation}
G^{(d)} = \left( \begin{matrix}  \frac{2}{d} & \dots &  \frac{2}{d} \\  \vdots & \ddots & \vdots \\  \frac{2}{d} & \dots &  \frac{2}{d} \end{matrix} \right ) - I_{d},
\label{2dcoingen}
\end{equation}
where $I_{d}$ is the identity matrix and $d$ is the size of the coin.
In the case of a square lattice, $d = 4$ for the four choices of direction
at each lattice site, and eq.~(\ref{2dcoingen}) reduces to
\begin{equation}
G^{(4)} = \frac{1}{2}\left( \begin{matrix} -1 & 1 & 1 & 1 \\ 1 & -1 & 1 & 1 \\ 1 & 1 & -1 & 1 \\ 1 & 1 & 1 & -1 \end{matrix} \right ).
\label{2dcoin}
\end{equation}
We also need a different coin for the marked state.  As we show later,
it is optimal to invert the phase of the $G^{(4)}$ coin operator,
\begin{equation}
G^{(4)}_m = \frac{1}{2}\left( \begin{matrix} 1 & -1 & -1 & -1 \\ -1 & 1 & -1 & -1 \\ -1 & -1 & 1 & -1 \\ -1 & -1 & -1 & 1 \end{matrix} \right ).
\label{markedcoin2d}
\end{equation}
To carry out the quantum walk search algorithm, we start with the quantum
walker in an equal superposition of all the possible sites in the lattice,
and the coin in an equal superposition of all directions,  
\begin{equation}
|\psi(0)\rangle = \frac{1}{\sqrt{4N}}\sum_{x=1}^N\sum_{c=0}^3 |x,c\rangle.
\end{equation}
We use periodic boundary conditions at the edges of the
$\sqrt{N}\times\sqrt{N}$ square lattice.  

\begin{figure}[!tb]
    \centering
	\includegraphics[width=0.45\textwidth]{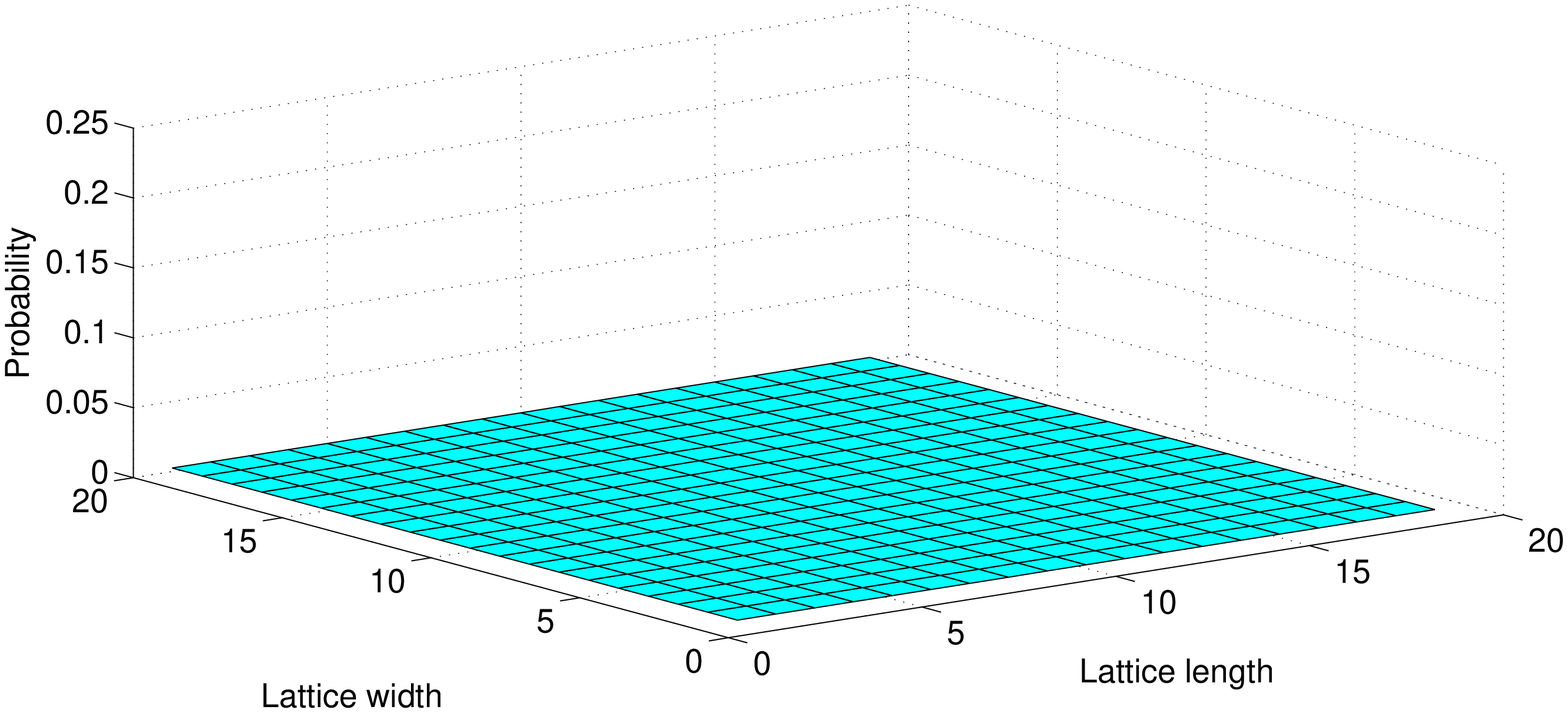}
	\includegraphics[width=0.45\textwidth]{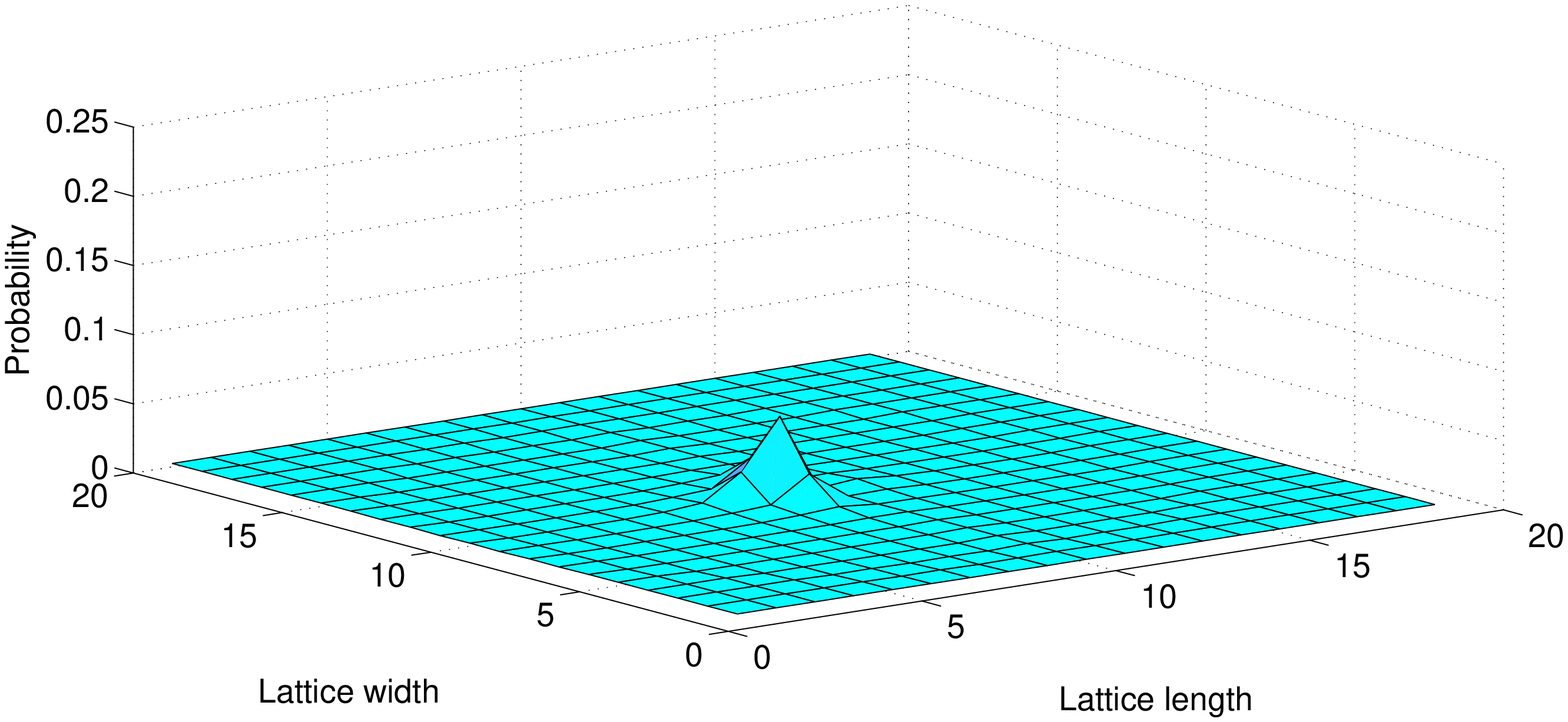}
	\includegraphics[width=0.45\textwidth]{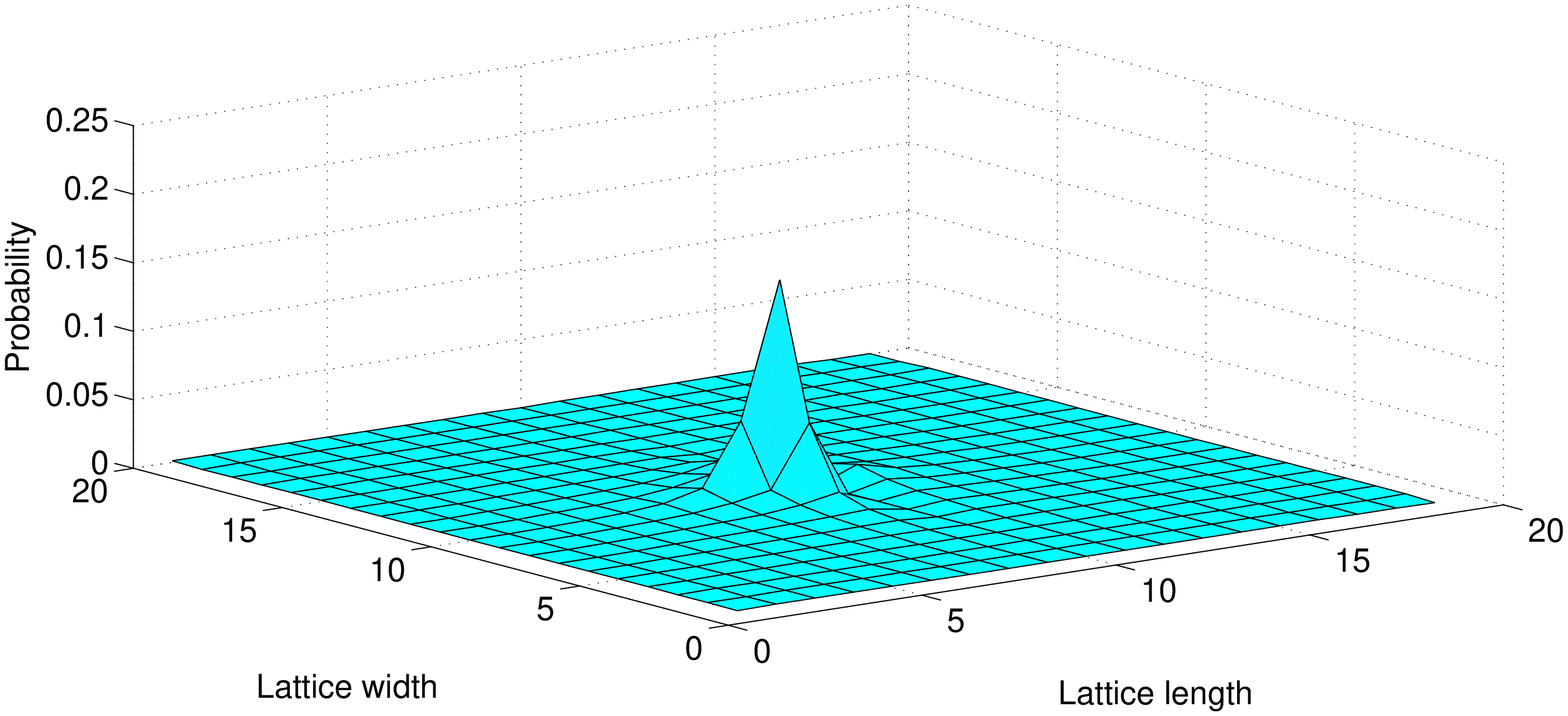}
	\includegraphics[width=0.45\textwidth]{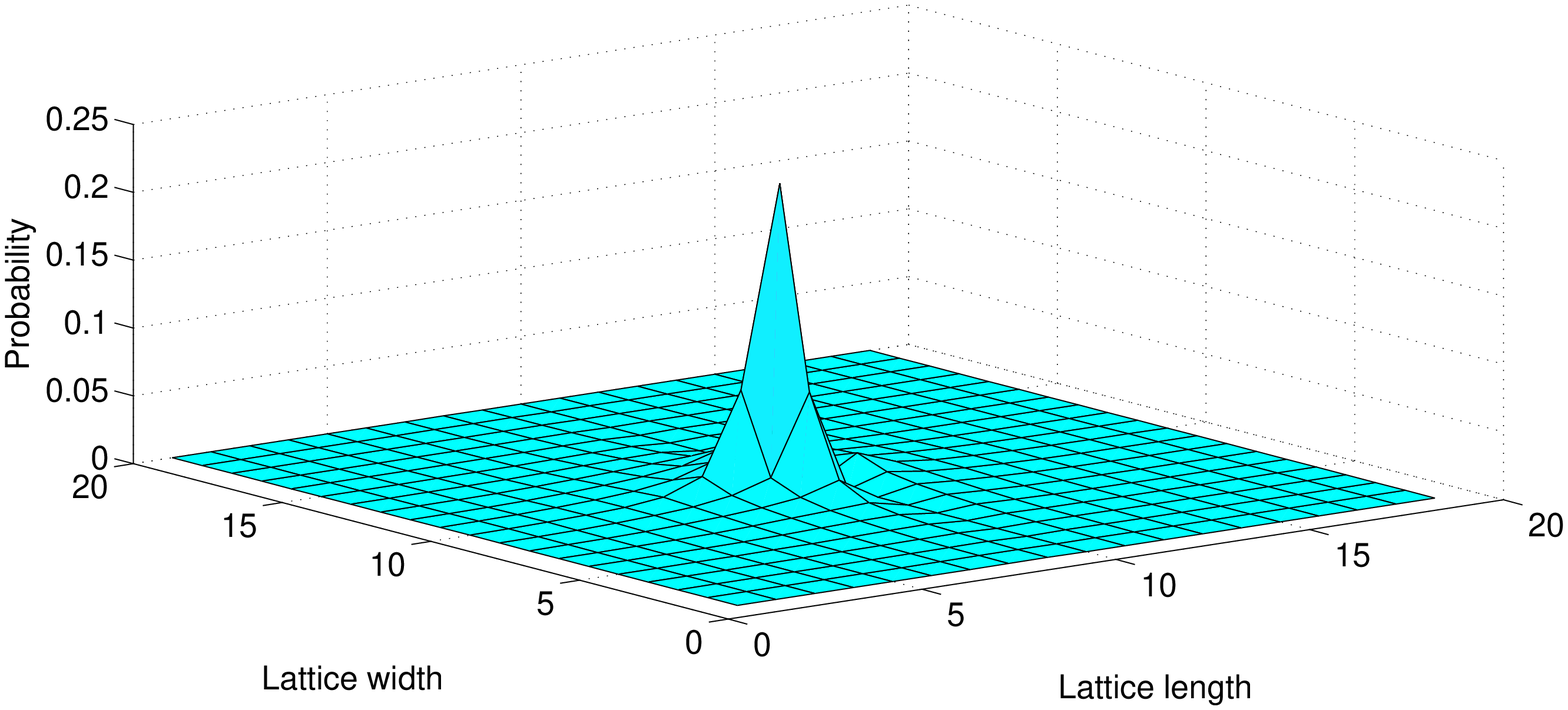}
	\caption{Probability distribution of a discrete time quantum walk
	search on 400 vertices arranged in a $20\times 20$ square
	with periodic boundary conditions, evolved for 0, 10, 20 and 32
	timesteps. The marked vertex is at position 190.}
    \label{twod4set}
\end{figure}

\begin{figure}[!tb]
    \begin{minipage}{\columnwidth}
	\centering
	\includegraphics[width=0.65\textwidth]{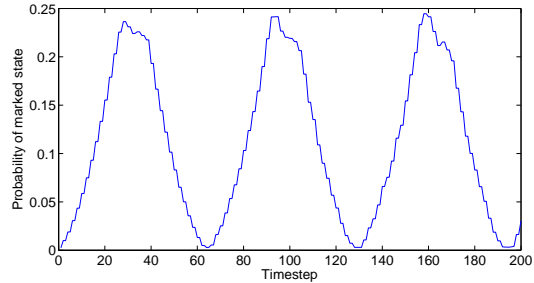}
	\caption{Probability of marked state over 200 timesteps
	on a $20\times 20$ grid with periodic boundary conditions.
	The marked vertex is at position 190.}
	\label{maxprob2d}
    \end{minipage}
\end{figure}

Figure~\ref{twod4set} shows how the distribution of the
walker evolves with time for a $20\times 20$ lattice, i.e. $N=400$.
We are interested in how quickly the quantum walker finds the marked state:
fig.~\ref{maxprob2d} shows the probability of being at
the marked state for each timestep as the walk proceeds. 
\begin{figure}[h!bt]
    \begin{minipage}{\columnwidth}
	\centering
	\includegraphics[width=0.55\textwidth]{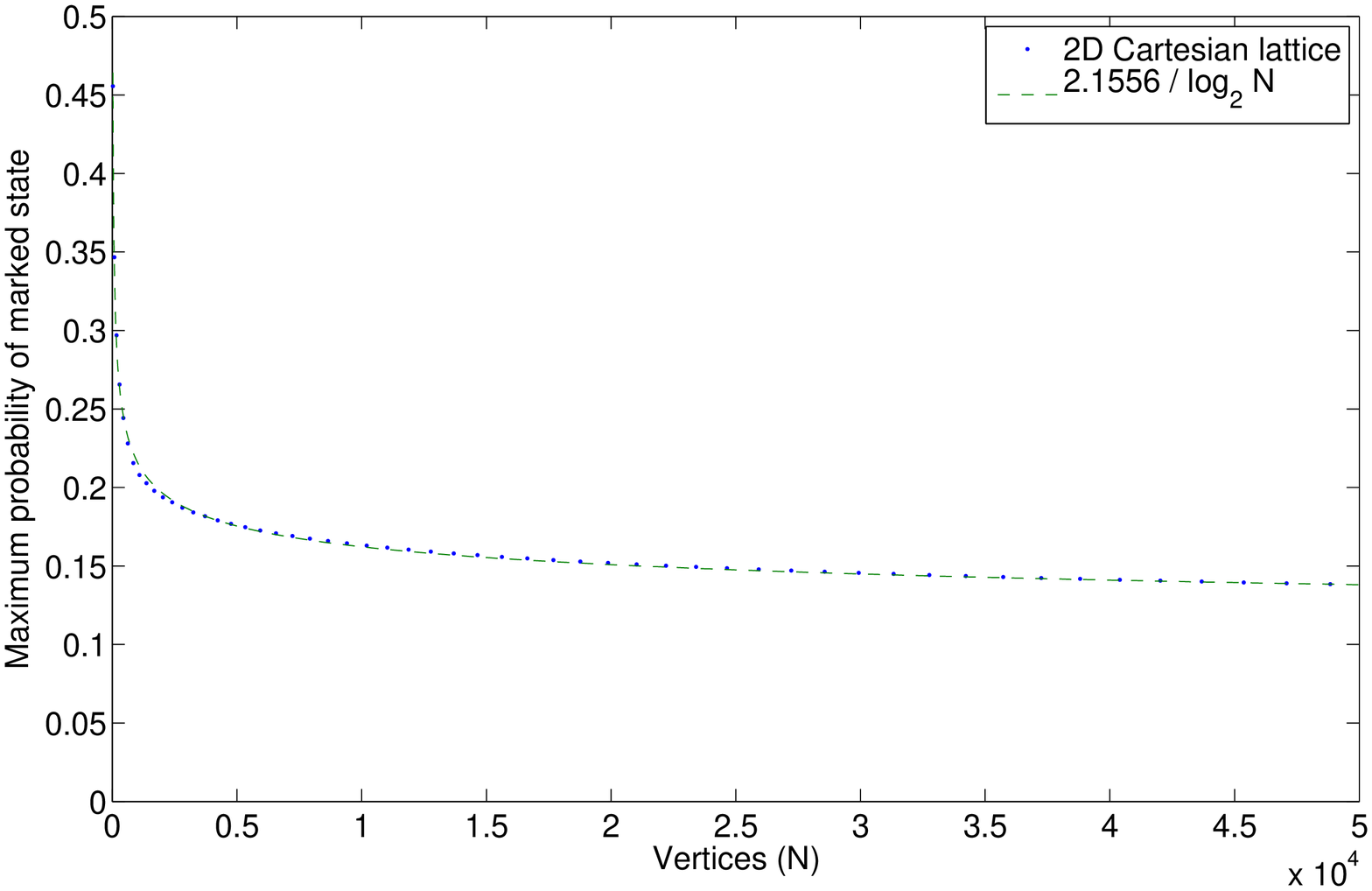}
	\caption{Maximum of the first peak in the probability
	of being at the marked state for
	different sized data sets, using the optimal marked state coin
	in eq.~(\ref{markedcoin2d}) on a 2D lattice of size
	$\sqrt{N}\times\sqrt{N}$, plotted against $N$
	(circles).  Also shown is the closest fit to our data, $2.16 / \log_{2} N$ (dashes).}
	\label{maxprob2dsizea}
    \end{minipage}
\end{figure}
It has periodic behaviour with the first peak occuring at roughly 
$t=(\pi/2)\sqrt{N} \simeq 32$.
The maximum probability for $N=400$ is around 0.23.
This can be increased as close to 1 as desired by standard
amplification techniques (repeating the search a few times).
Subsequent peaks occur somewhat later than $t=2(\pi/2)\sqrt{N}$ etc.,
but we are interested in the oscillatory behaviour not for the precise
timings of the peaks, but rather for the necessity of knowing when the first
peak occurs in order to measure the walker's position at the optimum time.
In fact, as can be seen in fig.~\ref{maxprob2d}, the peaks are quite
broad, so even if an error occurs in when to measure, it only means a lower
probability of finding the marked state, this is only a constant
extra overhead on the amplification. For example, if the state of the walker was measured at half the optimal number of timesteps ($t=(\pi/4)\sqrt{N} \simeq 16$), the probability of the walker being measured in the marked state is roughly half that of the maximum possible ($p\approx0.1$). 
The maximum probability also varies with the size of the data set,
the theoretical value of $O(1/\log_2 N)$ from
\citeauthor{ambainis04b}~\cite{ambainis04b} is numerically confirmed in our
results in fig.~\ref{maxprob2dsizea} with a small pre-factor of just over 2.

To explore how the coin affects the search result, we can
introduce a phase into the marked state coin operator, eq.~(\ref{markedcoin2d}),
\begin{equation}
G^{(4)}_{\phi, m} = e^{i\phi}G_{m}^{(4)},
\label{phase4dcoin}
\end{equation}
where $0\le\phi\le\pi$. The standard $G^{(4)}$ coin operator, eq.~(\ref{2dcoin}), corresponds
to $\phi=0$, and the marked coin operator used before, $G^{(4)}_m$,
eq.~(\ref{markedcoin2d}), corresponds to
$\phi=\pi$. A bias can also be introduced by a matrix of the form,
\begin{equation}
G^{(4)}_{\delta} = \left( \begin{matrix}  \delta & a+ib & a+ib & a+ib \\ a+ib & \delta & a+ib & a+ib \\ a+ib & a+ib & \delta & a+ib\\ a+ ib & a+ib & a+ib & \delta \end{matrix} \right ),
\label{bias2dcoin}
\end{equation}
where $0.5\le\delta\le1$. We solve this to give expressions for $a$ and $b$ in terms of the bias parameter $\delta$ in order to interpolate between the identity and marked state $G_{m}^{(4)}$ matrix, eq.~(\ref{markedcoin2d}). Taking $\delta=1$ makes the marked state coin into the
identity operator, while $\delta=0.5$ corresponds to the $G^{(4)}_m$ operator.
\begin{figure}[!bt]
\begin{minipage}{\columnwidth}
\centering
\includegraphics[width=0.65\textwidth]{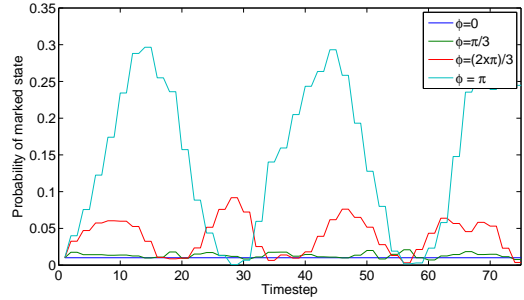}
\caption{Probability of marked state over 75 timesteps
for $N=100$ i.e., a $10\times 10$ grid with marked state at 45,
using the marked state coin in eq.~(\ref{phase4dcoin})
with $\phi=0$, $\pi/3$, $2\pi/3$ and $\pi$.}
\label{maxprob2dphase}
\end{minipage}
\end{figure}
\begin{figure}[!bt]
\begin{minipage}{\columnwidth}
\centering
\includegraphics[width=0.65\textwidth]{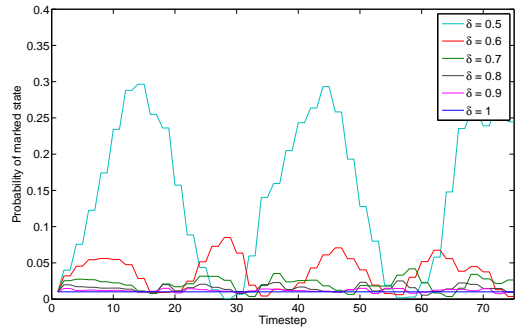}
\caption{Probability of marked state over 75 timesteps
for $N=100$ i.e., a $10\times 10$ grid with marked state at 45,
using the marked state coin in eq.~(\ref{bias2dcoin})
with $\delta=0.5$, $0.6$, $0.7$, $0.8$, $0.9$ and $1$.}
\label{maxprob2dbias}
\end{minipage}
\end{figure}
These variations have been chosen
to preserve the symmetry of the coin operator.
Figures \ref{maxprob2dphase} and \ref{maxprob2dbias} show the effect of varying both the phase $\phi$ and the bias $\delta$ showing the largest probability of finding the marked state is for $\phi=\pi$ and $\delta=0.5$, justifying our choice of
the optimal marked state coin operator.

\begin{figure}[!tb]
    \centering
	\includegraphics[width=0.55\textwidth]{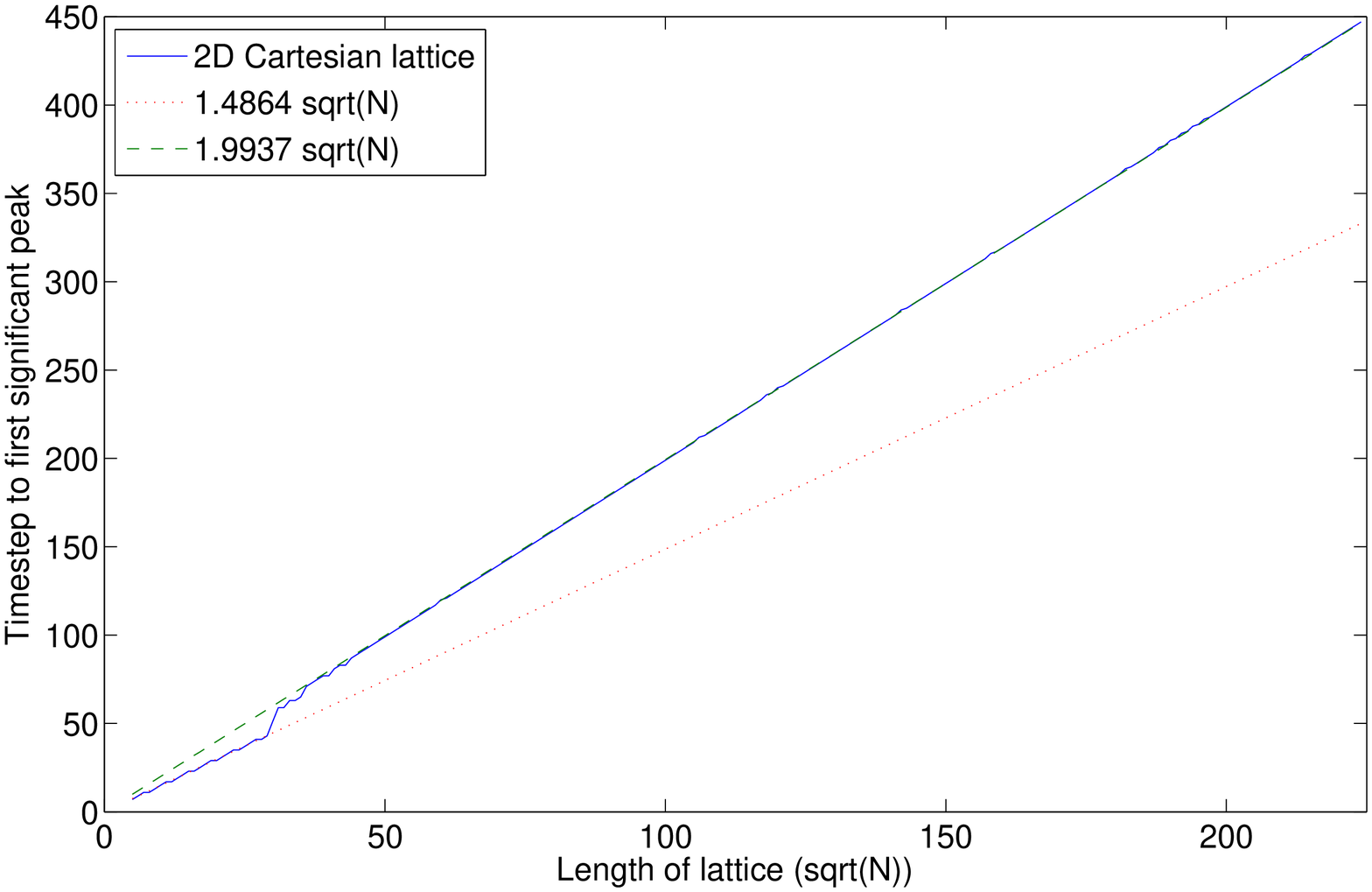}
	\caption{Time step at which the first peak in the probability
	of being at the marked state occurs for different sized data sets,
	using the optimal marked state coin in eq.~(\ref{markedcoin2d}),
	plotted against $\sqrt{N}$. Also shown are closest fits for $\sqrt{N} < 32$ (dotted) and for $\sqrt{N} > 32$ (dashes).}
    \label{maxprob2dtime}
\end{figure}

Our numerical results are shown in fig.~\ref{maxprob2dtime}, where we see the closest fit for the data up to $N=32^2$ is $1.49\sqrt{N}$, but there is a `kink' near $N=32$ that pushes it up to $1.99\sqrt{N}$.
 Our numerical evidence thus supports a scaling $O(\sqrt{N})$. As the maximum probability of finding the marked state scales as $O(1/\log_2 N)$, we must use amplitude amplification techniques to bring this up to $O(1)$. This requires $O(\sqrt{\log N})$ repetitions \cite{brassard02a}. The total scaling we find with our numerical simulations is thus $O(\sqrt{N \log N})$, which is consistent with the expected optimal scaling for two dimensions proved by \citeauthor{magniez08a} \cite{magniez08a}. \citeauthor{tulsi08a} \cite{tulsi08a} gives an algorithm that also meets this optimal scaling, but uses 
ancillas, while our simulations do not, they follow the method of \citeauthor{shenvi02a} \cite{shenvi02a} using a standard coined quantum 
walk.

\section{Quantum walk search on the line}
\label{sec:1Dsearch}

We now examine how the quantum walk search algorithm behaves for data
arranged on a line. \citeauthor{szegedy04a} \cite{szegedy04a} previously proved that a quantum walk search approach can only find a marked state in 
time $O(N)$ with probability $1/N$. We confirm this numerically considering both a line segment, and a loop (cycle)
where periodic boundary conditions are applied at the ends.
The loop is less physical (for most tape storage,
the ends are far apart from each other),
but allows us to investigate the behaviour of the algorithm without the
edge effects the ends introduce.
For the line segment, we use a reflecting boundary condition.
This means we have to use a different coin at the edges, which has the effect
of introducing two spurious marked states.

Since symmetry is important in the quantum walk search algorithm in 
higher dimensions, we also investigate a more symmetric version of 
the Hadamard operator,
\begin{equation}
H^{(\text{sym})}_{\delta} = \left( \begin{matrix} \sqrt{\delta} & i\sqrt{1-\delta} \\ i\sqrt{1-\delta} & \sqrt{\delta} \end{matrix} \right ).
\label{complexcoindelta}
\end{equation}
For $\delta=0.5$, this reduces to
\begin{equation}
H^{(\text{sym})} = \frac{1}{\sqrt{2}}\left( \begin{matrix} 1 & i \\ i & 1 \end{matrix} \right ).
\label{complexcoin}
\end{equation}
We vary $\delta$ from zero to one, to see how this affects the performance.
The initial state for the quantum walk algorithm on the line needs to 
match the symmetry of the chosen coin operator.  For the Hadamard, we use
\begin{equation}
|\phi(0)\rangle = \frac{1}{\sqrt{N}}\sum_{x=0}^N \frac{1}{\sqrt{2}}(|x,0\rangle +i|x,1\rangle),
\end{equation}
and for the symmetric coin operator we use
\begin{equation}
|\phi(0)\rangle = \frac{1}{\sqrt{N}}\sum_{x=0}^N \frac{1}{\sqrt{2}}(|x,0\rangle +|x,1\rangle).
\end{equation}

\begin{figure}[!tb]
    \centering
	\includegraphics[width=0.45\textwidth]{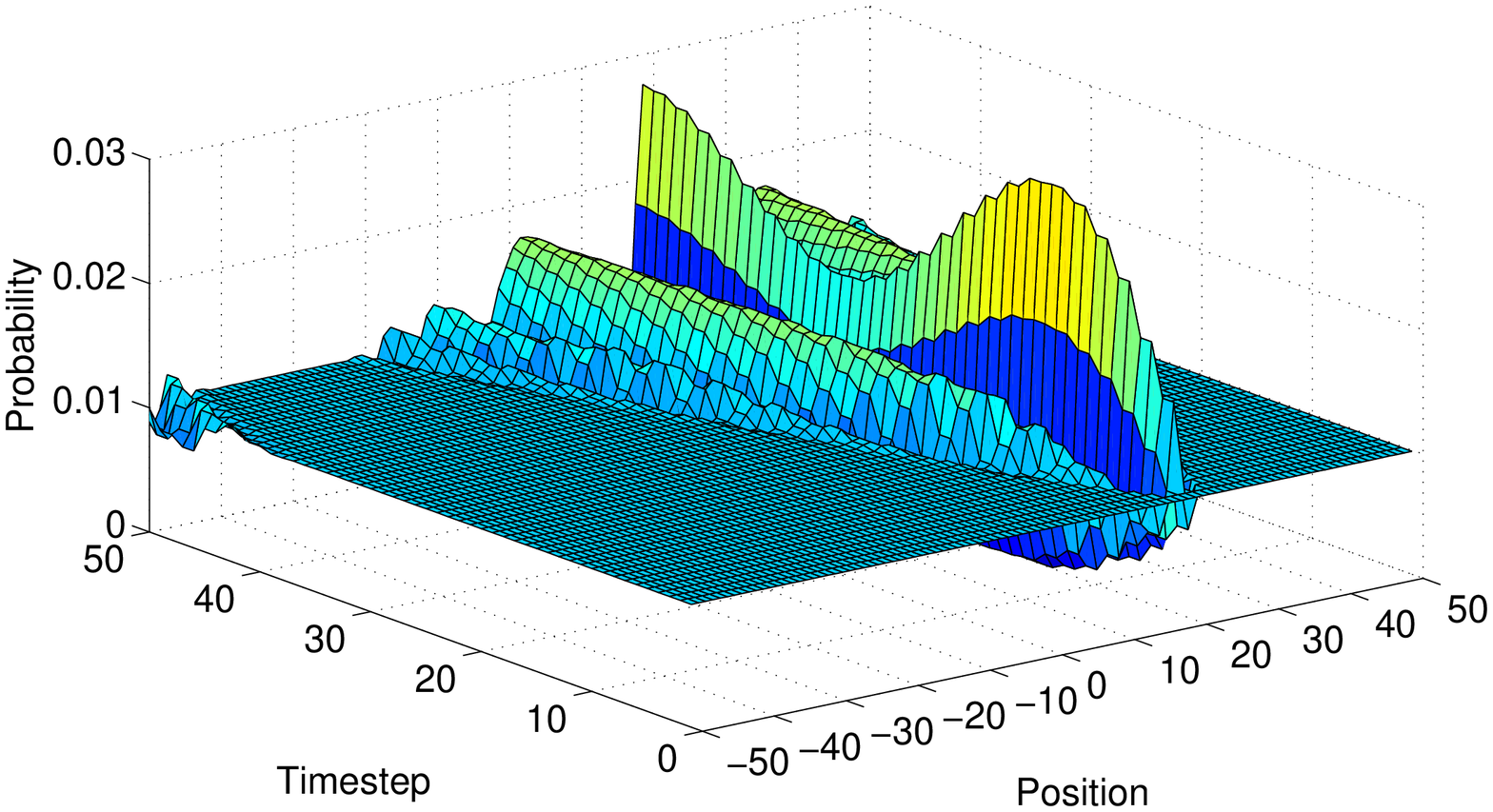}
	\includegraphics[width=0.45\textwidth]{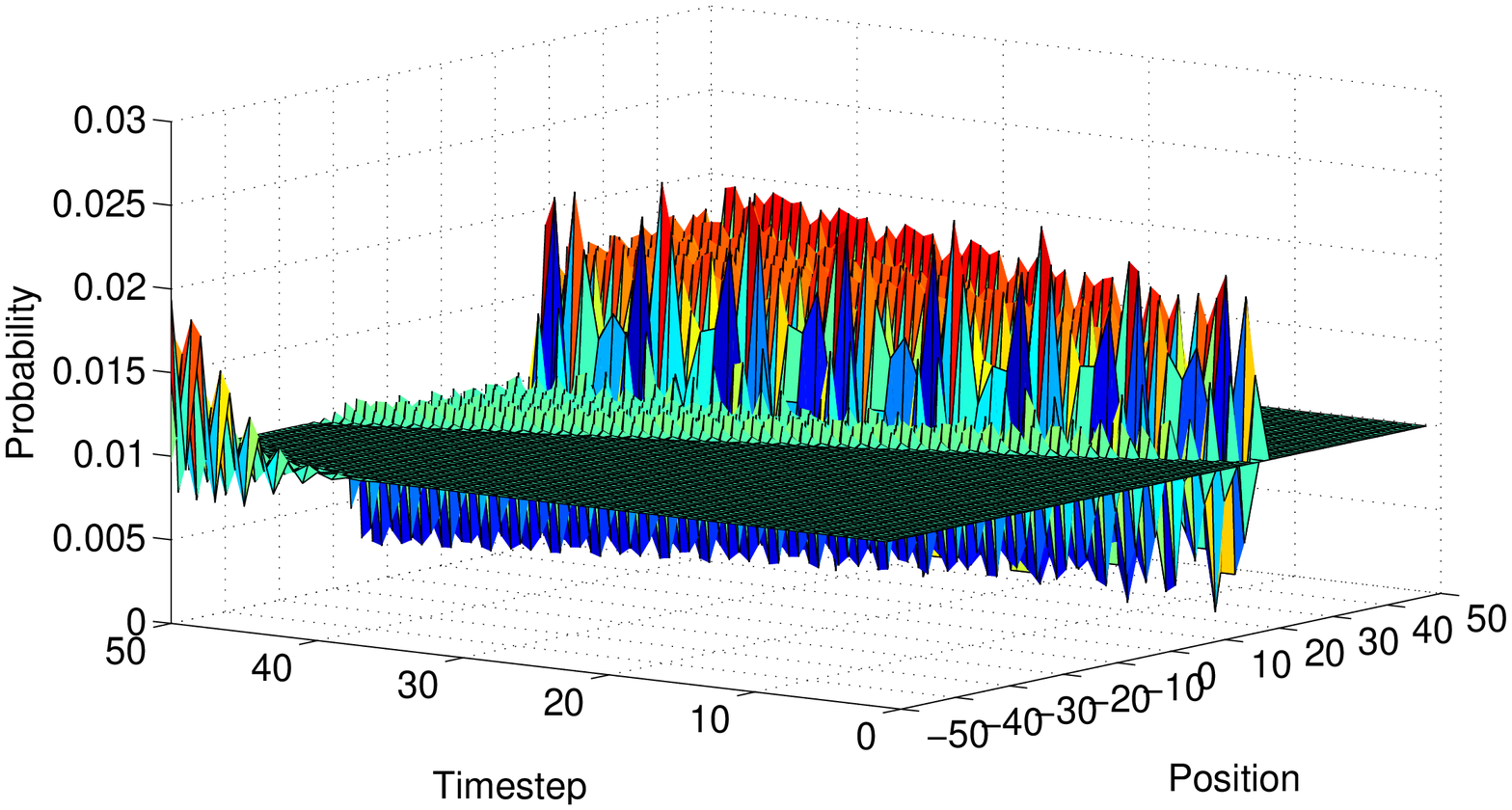}
	\includegraphics[width=0.45\textwidth]{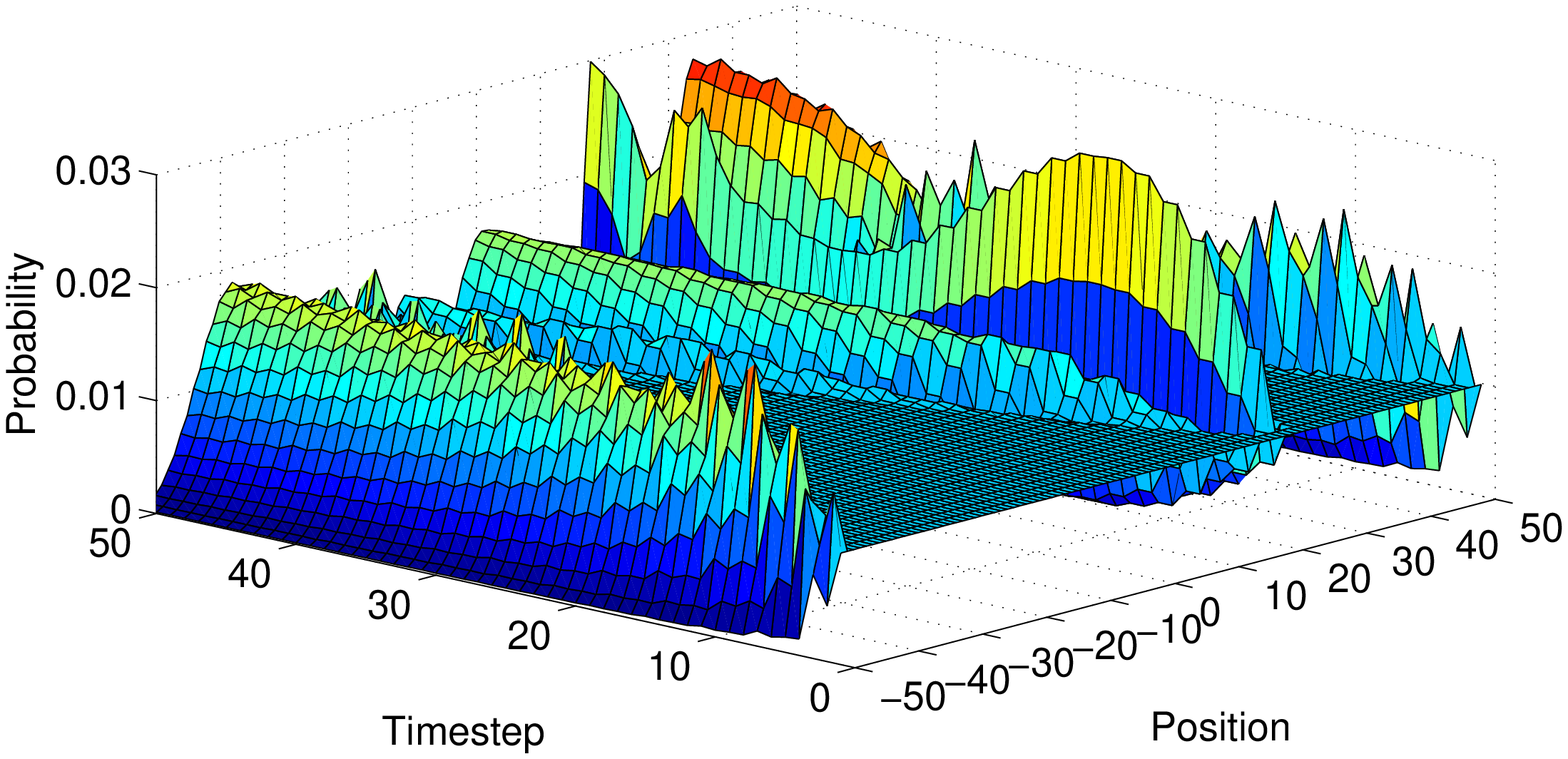}
	\includegraphics[width=0.45\textwidth]{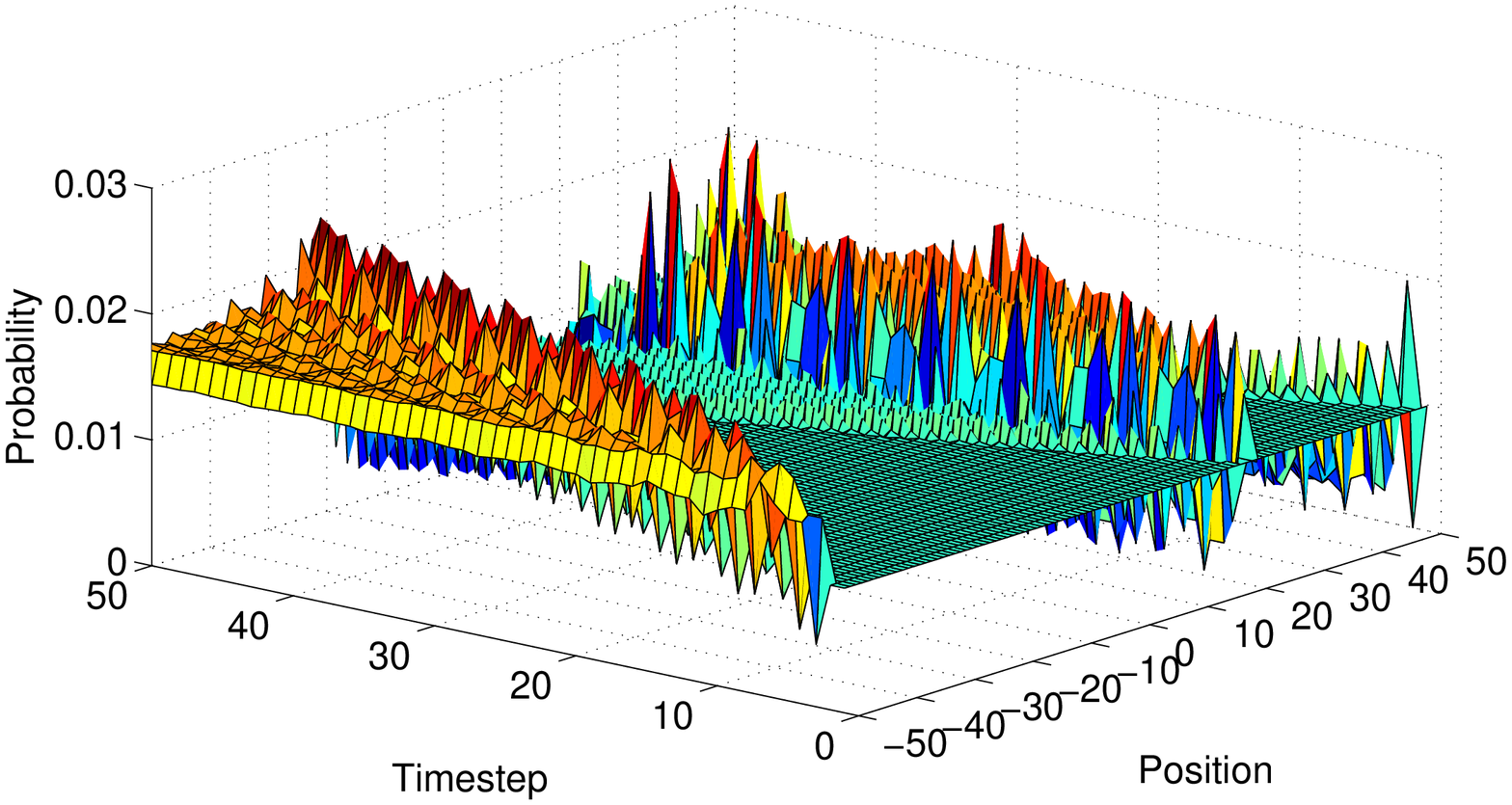}
	\caption{Probability distribution of the quantum walk search 
	of $N=101$ items arranged on a line, after 50 time steps
	with marked state at position 20, for a
	symmetric coin and periodic boundary conditions (top left),
	Hadamard and periodic (top right),
	symmetric and reflecting (bottom left), and
	Hadamard and reflecting (bottom right).}
    \label{1dwalk}
\end{figure}

First we contrast the symmetric coin operator, eq.~(\ref{complexcoin}) with
the standard Hadamard operator, eq.~(\ref{hadamard}). In fig.~\ref{1dwalk} we see the contrast between the two possible boundary conditions (periodic and reflecting). We show each condition with both the coin operators described previously, eqs.~(\ref{hadamard}) and (\ref{complexcoin}), for all dynamics (negating the phase for the marked state operator).
We find that the symmetric coin operator gives a
more smoothly varying probability distribution about the marked
state and we can see oscillations in the probability
of finding the marked state.
Superficially, these results look similar to the square lattice.
However, the square lattice has a peak probability at the marked state of
around 0.3 for a $10\times 10$ square lattice with $N=100$.
On a line with $N=101$, the peak in the probability is only around 0.028,
This is not significantly larger than the uniform distribution,
which has a probability of 0.01 for any site.
The Hadamard coin operator varies over a period of around seven time steps,
regardless of the other parameters, and shows slightly higher probability
spreading out from the marked state in two soliton-like waves.  This
supports the case for symmetry being important for quantum walk searching.
Reflecting boundary conditions also produce spreading soliton-like waves from 
the boundaries for both symmetric and Hadamard coins.

\begin{figure}[!tb]
    \centering
	\includegraphics[width=0.45\textwidth]{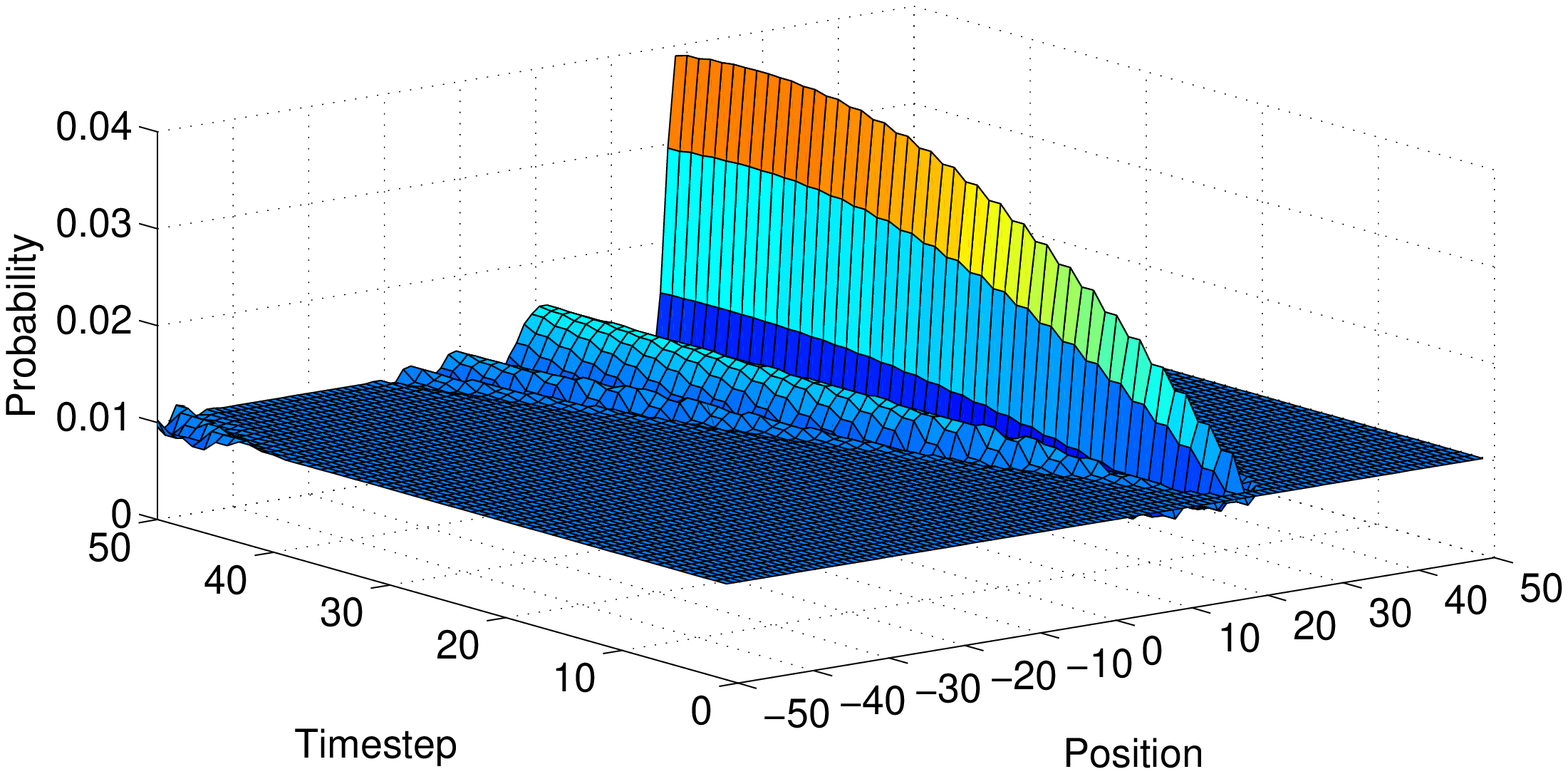}
	\includegraphics[width=0.45\textwidth]{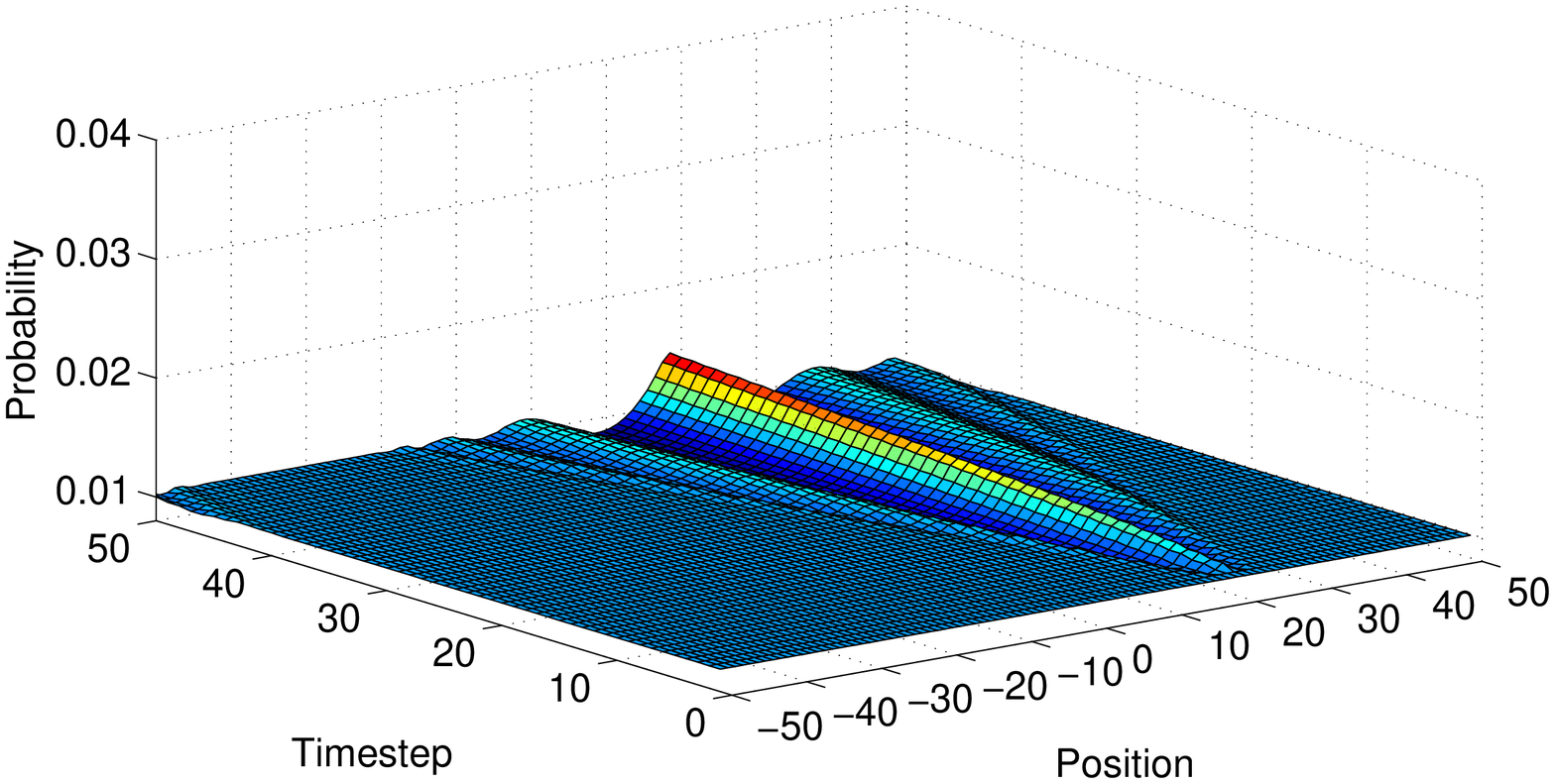}
	\includegraphics[width=0.45\textwidth]{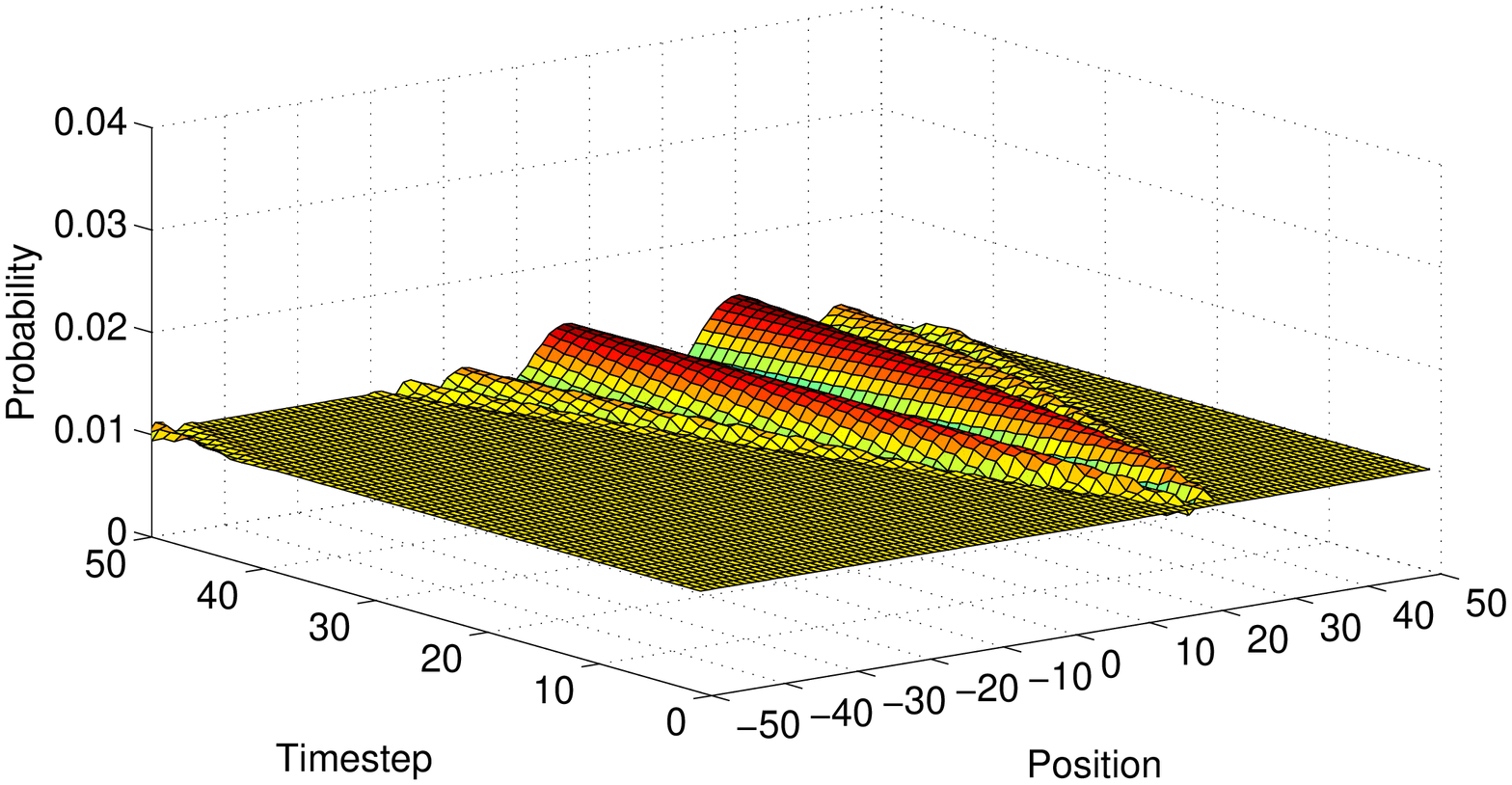}
	\includegraphics[width=0.45\textwidth]{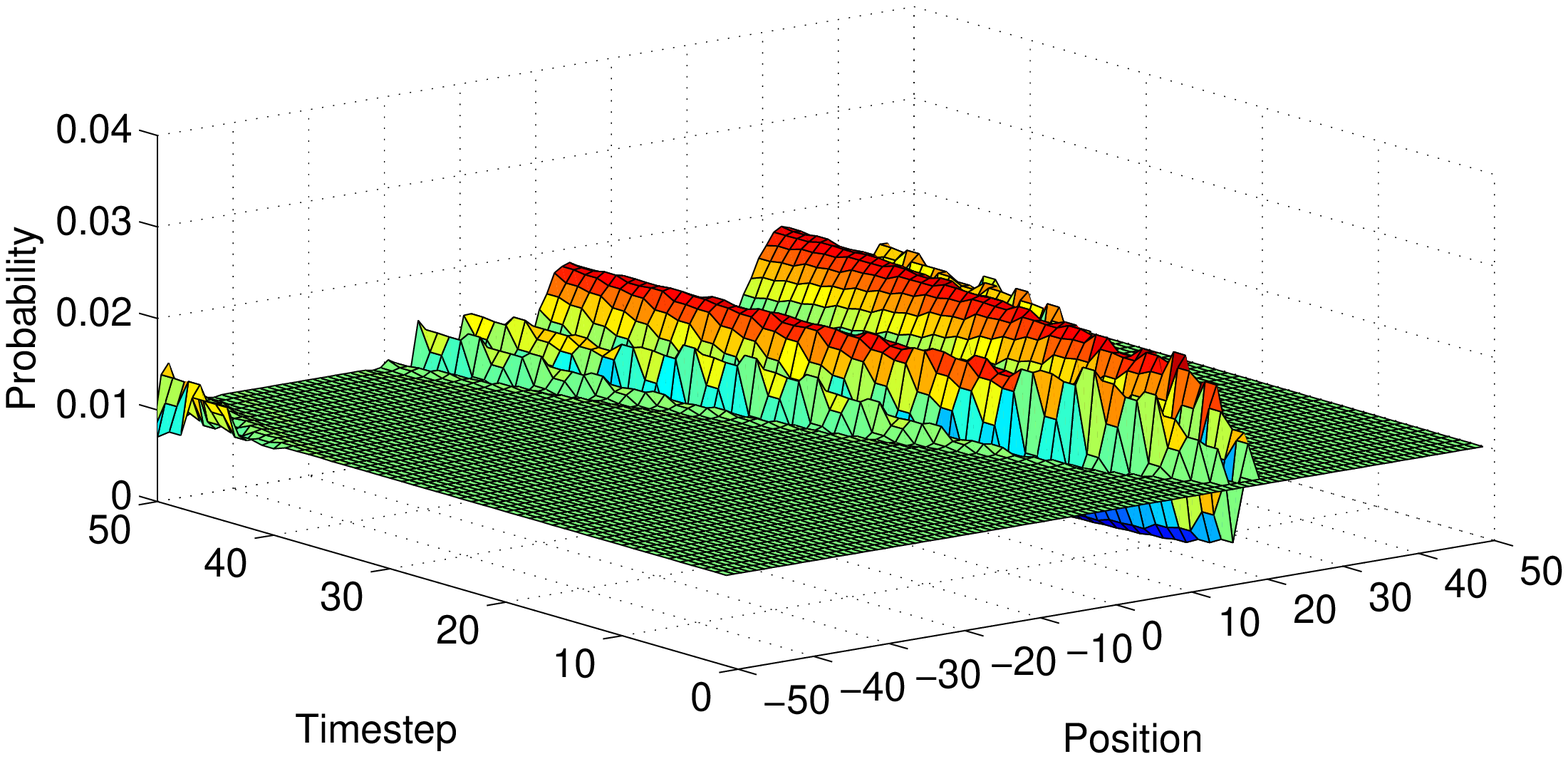}
	\caption{Probability distribution of the quantum walk search 
	of $N=101$ items arranged on a line, after 50 time steps
	with marked state at position 20, for a
	symmetric coin and periodic boundary conditions with the marked
	coin operator given by eq.~(\ref{complexcoindelta}) with
	$\delta=0.15$ (top left),
	$\delta=0.45$ (top right),
	$\delta=0.65$, (bottom left) and
	$\delta=1$ (bottom right).}
    \label{oscill}
\end{figure}

\begin{figure}[!tb]
    \centering
	\includegraphics[width=0.75\textwidth]{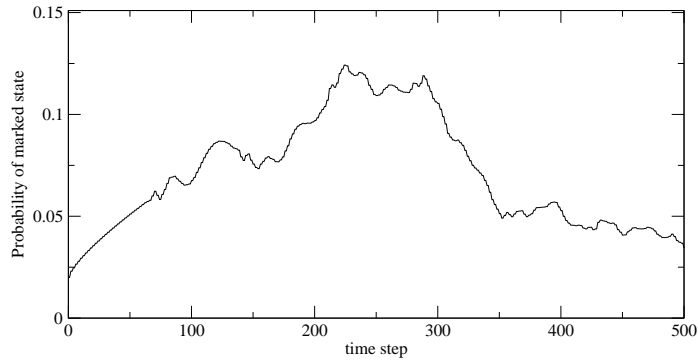}
	\caption{Probability of the marked state for each time step
	for a line of $N=50$ sites with periodic boundary conditions
	and $\delta=0.45$, run for 500 time steps.}
    \label{linelongpmarked}
\end{figure}

Concentrating on the case of a symmetric coin operator and
periodic boundary conditions, we now consider what happens when
$\delta$ is varied.  For $\delta$ increasing from zero to a half,
the period of the oscillations increases towards infinity as
$\delta\rightarrow 0.5$, the value for which the marked coin operator
becomes the same as the unmarked coin operator and the distribution
remains uniform.
Increasing $\delta$ above 0.5, we find that instead of
finding the marked state, in effect it ``un-finds''
it with the probability of being in the marked state decreasing below
the uniform distribution.
This is due to the fact the coin is now biased in the wrong direction
and so is giving away more and more probability instead of retaining it.
Figure~\ref{oscill} shows the variation in probability distribution
with $\delta$, compare with $\delta=0$ from fig.~\ref{1dwalk} (top left).
For the symmetric quantum walk with periodic boundary conditions,
evolving for longer times with $\delta\simeq 0.45$
eventually results in a peak in the probability of the marked state
approaching $2\pi/N$, but only after around $5N$ or more time steps,
see fig.~\ref{linelongpmarked}.
This is obviously not useful, either in the size of the peak probability,
which we would like to scale with at least $\log N$ as it does for the search
in two dimensions, nor with the number of time steps, which far exceeds
the classical worst case of $N$.

The quantum walk search algorithm used for data on a line is thus completely
ineffective for the parameters we have considered:
it does not find the marked state with significant probability even when run
for as long as the worst case classical time of $N$ steps for $N$ items.
Of course, we can easily specify a quantum version of the classical
algorithm that does find the marked state in $N$ steps.
For example, start in the state $|0,1\rangle$ and use the identity as the 
coin operator everywhere except the marked state.  This causes the
walk to hop deterministically along the line.  At the marked state, use
$\sigma_x$ for the coin operator.  This will flip the coin from $|1\rangle$
to $|0\rangle$ and thus reverse the direction of the
walker.  If the position of the walker is measured after $N$ steps, the
current location allows you to work out where it turned round,
and thus locate the marked state.  This method uses only a single measurement.
If you allow measurements at every step, then of course you can
immediately find out if the walker has arrived at the marked state
by testing the state of the coin. 

A classical random walk searching algorithm on a line, with equal probability of moving left and right, would take $O(N^2)$ to find a marked item. Using the techniques by \citeauthor{magniez08a} \cite{magniez08a} and \citeauthor{krovi10a} \cite{krovi10a}, a quantum analogue of this classical walk can be defined which would give a quadratic speed up to $O(N)$. This would give a constant probability, $O(1)$, for the maximum probability of the marked state. However, as shown by \citeauthor{szegedy04a} \cite{szegedy04a}, the value of this maximum probability at the marked state would be $1/N$ thus rendering the algorithm ineffective.

\section{Two-dimensional structures}
\label{sec:2dstructures}

In order to test how the arrangement of the data affects the search algorithm for a fixed spatial dimension, we investigated its efficiency on other regular structures, namely, the hexagonal lattice, the 2D Cartesian lattice with diagonal links and the Bethe lattice of degree three. These structures have very different connectivity between data points (vertices). The Bethe lattice has only one shortest path to the marked state due to its `branching' structure, whereas the 2D Cartesian lattice with diagonal links has, in general, many shortest paths.

\subsection{Hexagonal lattice}

\begin{figure}[!tb]
    \centering
	\includegraphics[width=0.45\columnwidth]{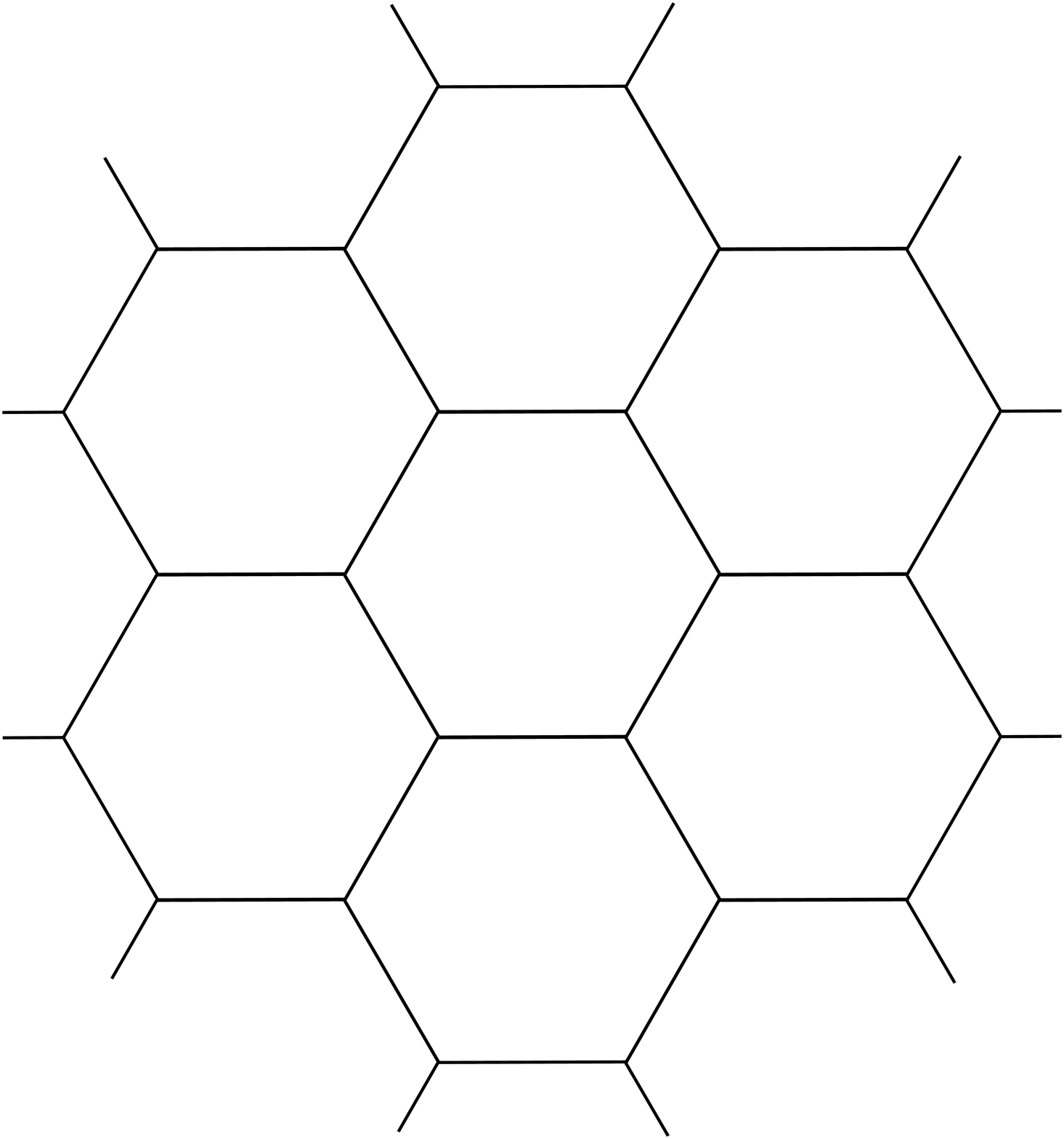}
	\includegraphics[width=0.45\columnwidth]{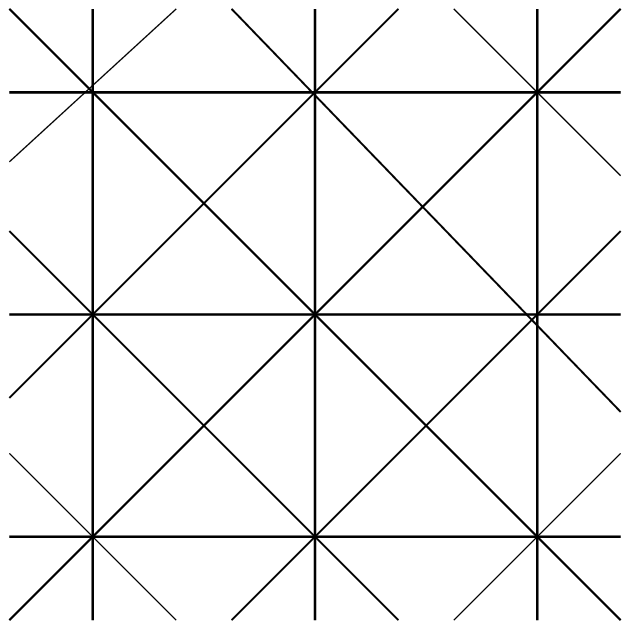}
	\caption{A portion of the structures we discuss in sec.~\ref{sec:2dstructures}, on which we study the quantum walk search algorithm. We impose periodic boundary conditions on both structures. Left: Hexagonal lattice, each vertex is of degree $d = 3$. Right: 2D Cartesian lattice with diagonal links, each vertex is of degree $d=8$.}
    \label{connectstruct}
\end{figure}

We study a hexagonal lattice, where each vertex is of degree $d = 3$, with periodic boundary conditions as shown in fig.~\ref{connectstruct} (left).
We use the Grover coin, eq.~(\ref{2dcoingen}), in dimension three,
\begin{equation}
G^{(3)} = \frac{1}{3}\left( \begin{matrix} -1 & 2 & 2 \\ 2 & -1 & 2  \\ 2 & 2 & -1  \end{matrix} \right ),
\label{hexcoin}
\end{equation}
and use the inverse of this for the marked state in the same way as for the 2D Cartesian lattice described in section \ref{sec:2Dsearch},

\begin{equation}
G^{(3)}_{m} = \frac{1}{3}\left( \begin{matrix} 1 & -2 & -2 \\ -2 & 1 & -2  \\ -2 & -2 & 1  \end{matrix} \right ).
\label{hexcoinm}
\end{equation}

\begin{figure}[!tb]
    \centering
	\includegraphics[width=0.65\textwidth]{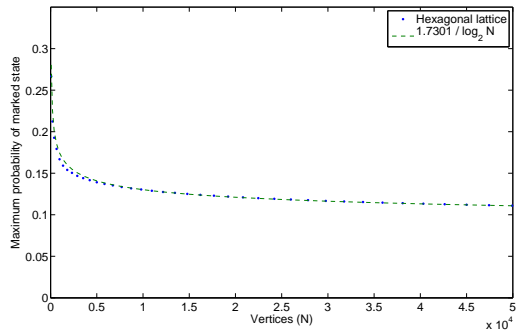}
	\caption{Maximum of the first peak in the probability of being at the marked state for different sized data sets for a hexagonal lattice, plotted against N (circles). Also shown is the closest fit to our data, $1.73 / \log_{2} N$ (dashes).}
    \label{hexprob}
\end{figure}

\begin{figure}[!htb]
    \centering
	\includegraphics[width=0.65\textwidth]{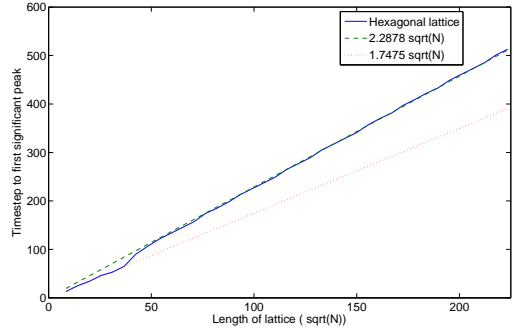}
	\caption{Time step at which the first significant peak in the probability of being at the marked state occurs for different sized data sets plotted against $\sqrt{N}$, for a hexagonal lattice. Also shown are closest fits for $\sqrt{N} < 42$ (dotted) and for $\sqrt{N} > 42$ (dashes).}
    \label{hextime}
\end{figure}

We can see how the search algorithm performs on the hexagonal lattice in figs.~\ref{hexprob} and \ref{hextime}. The maximum probability of the marked state is lower than that of the more highly connected 2D Cartesian lattice. We still find the $O(1/\log_{2} N)$ scaling with the size of the data set but with a smaller pre-factor, 1.73, compared with just over 2 for the 2D Cartesian lattice. The first significant peak in the probability of the marked state occurs later than for the 2D Cartesian lattice. This shows that not only is the algorithm dimensionally dependent, but it also depends on the actual connectivity of the structure. Interestingly, we also find the unusual `kink' we saw in the 2D Cartesian case but at a different point. The time to find the marked state still best fits a scaling of $O(\sqrt{N})$, still giving a algorithmic complexity of $O(\sqrt{N \log N})$, but with larger pre-factors before and after the `kink', 1.75 and 2.29 respectively compared to the prefactors of 1.49 and 1.99 before and after the kink in the 2D Cartesian lattice. The basic scaling (without prefactors) of $O(\sqrt{N \log N})$ matches recent analytical results by \citeauthor{abal10a} \cite{abal10a}. In the 2D Cartesian lattice, we found this jump in scaling when the size of the grid reached roughly $32^2$ vertices. For the hexagonal lattice we find this point is higher at roughly $42^2$ vertices in size.  

\subsection{Two-dimensional Cartesian lattice with diagonal links}

By adding diagonal links to the 2D Cartesian lattice, we create a more highly connected structure of degree $d=8$ at each vertex. This is shown in fig.~\ref{connectstruct} (right) and again periodic boundary conditions are imposed. The Grover coin, eq.~(\ref{2dcoingen}), is used which reduces to,

\begin{equation}
G^{(8)} = \frac{1}{4}\left( \begin{matrix} -3 & 1 & 1 & 1 & 1 & 1 & 1 & 1 \\ 1 & -3 & 1 & 1 & 1 & 1 & 1 & 1 \\ 1 & 1 & -3 & 1 & 1 & 1 & 1 & 1 \\ 1 & 1 & 1 & -3 & 1 & 1 & 1 & 1 \\ 1 & 1 & 1 & 1 & -3 & 1 & 1 & 1 \\ 1 & 1 & 1 & 1 & 1 & -3 & 1 & 1 \\ 1 & 1 & 1 & 1 & 1 & 1 & -3 & 1 \\ 1 & 1 & 1 & 1 & 1 & 1 & 1 & -3 \end{matrix} \right ),
\label{2d8coin}
\end{equation}
and the marked state operator as $G_{m}^{(8)} = - G^{(8)}$.

\begin{figure}[!tb]
    \centering
	\includegraphics[width=0.65\textwidth]{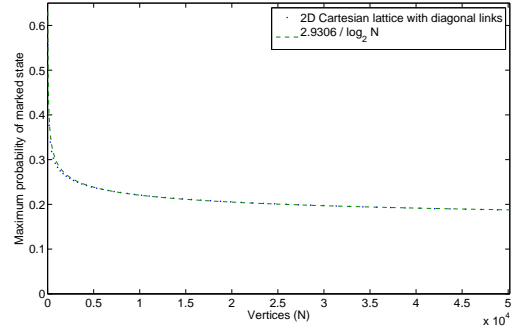}
	\caption{Maximum of the first peak in the probability of being at the marked state for different sized data sets on a 2D Cartesian lattice with diagonal links, plotted against N (circles). Also shown is the closest fit to our data, $2.93 / \log_{2} N$ (dashes).} 
    \label{2d8prob}
\end{figure}

\begin{figure}[!tb]
    \centering
	\includegraphics[width=0.65\textwidth]{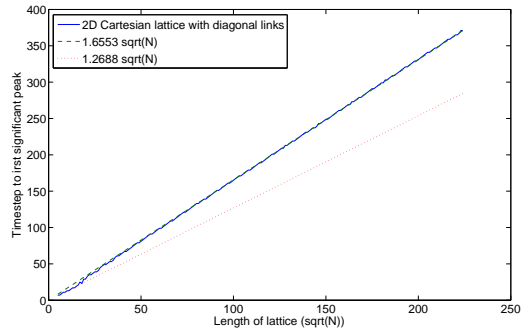}
	\caption{Time step at which the first significant peak in the probability of being at the marked state occurs for different sized data sets plotted against $\sqrt{N}$, for a 2D Cartesian lattice with diagonal links. Also shown are closest fits for $\sqrt{N} < 17$ (dotted) and for $\sqrt{N} > 17$ (dashes).}
    \label{2d8time}
\end{figure}

Figures \ref{2d8prob} and \ref{2d8time} show how the search algorithm performs on the 2D Cartesian lattice with diagonal links. We find an increase in the maximum probability of the marked state compared to the other two-dimensional structures. The basic 2D Cartesian lattice scales close to $2/\log_{2} N$, whereas, with the additional connectivity, the degree eight structure scales close to  $3/\log_{2}N$. The time to find the marked state is also faster. We notice the `kink' in scaling as in the other symmetrical structures but at a lower point in the graph. We see the algorithm still seems to scale close, $1.2688 \sqrt{N}$, to the optimal $\sqrt{N}$ for smaller structures up to roughly $17^2$ vertices in size. After that it jumps to scale as $1.6553 \sqrt{N}$ which is still closer to optimal than the basic 2D Cartesian lattice. Due to the scaling of the probability, the total algorithmic complexity is still the same as the hexagonal and 2D lattices, $O(\sqrt{N \log N})$. 

The search algorithm performs more efficiently here than either the hexagonal or the basic 2D Cartesian lattice. This is consistent with what we already noted for the hexagonal lattice, that the algorithmic efficiency scales with the connectivity of the structure. The connectivity here is much higher with every vertex being degree eight. This increases the number of paths the walker can take to the marked state. 

\subsection{Bethe lattice}

\begin{figure}[!tb]
    \centering
	\includegraphics[width=0.45\textwidth]{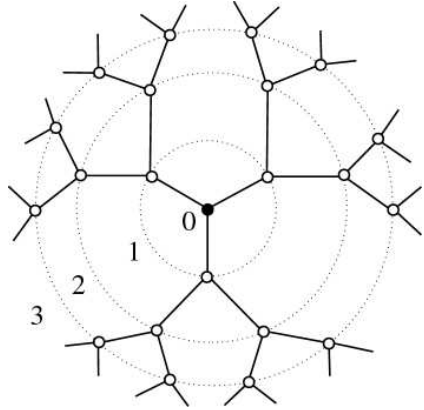}
	\caption{A segment of a Bethe lattice with fixed degree $d=3$. Three shells are shown here emanating from the central vertex.}
    \label{bethe3}
\end{figure}

The Bethe lattice (or Cayley tree) is a general structure which can have any fixed degree at all of its vertices. Its connectivity is somewhat different to the previous examples in that there are no loops in it, giving a tree-like structure with `branches' stemming from a vertex indefinitely. We work with a finite sized segment based on around a central vertex. A piece with vertices of degree three is shown in fig.~\ref{bethe3}. It can be seen from this example that the vertices on the branches form `shells' around the central vertex. The number of vertices in each shell is,

\begin{equation}
N_{s} = d(d-1)^{s-1}  \quad \text{where } s>0,
\end{equation}
where $N_{s}$ is the number of vertices in shell $s$ and $d$ is the degree of the vertices. The coin used in the degree three case of the Bethe lattice is the Grover coin of dimension three given by eq.~(\ref{hexcoin}). The marked state operator is just this same coin inverted as with the other structures, eq.~(\ref{hexcoinm}). We can't impose periodic boundary conditions on the Bethe lattice without creating loops in the structure. Instead, at the `ends' of the branches, we reflect the amplitude back upon itself. This is accomplished using the Grover coin at its limit, $d=2$, which is the $\sigma_{x}$ operation as shown in eq.~(\ref{reflectcoin}).

We find that if the marked state was present either at the central vertex or in the first shell then the algorithm is actually almost optimal in scaling, close to $O(\sqrt{N})$. The probability at this point is also high enough to allow the marked state to be distinguished from the remaining superposition. In fact, as the probability scales as $O(1)$, this is the total complexity and so would be optimal. This is an unrealistic scenario though and so the Bethe lattice would never, in general, be efficient for the search algorithm. Although we have only shown the degree three Bethe lattice here, we have also studied the Bethe lattice of degree four, and it performs in a similar fashion.

In contrast to the 2D Cartesian lattice, the position of the marked state in the Bethe lattice (which shell it occurs in) strongly affects the efficiency of the search algorithm. As the marked state moves away from the central vertex, the probability of the marked state is significantly lower than if the marked state is the central vertex. In fact, by the time the marked state is in the fourth shell, the probability of the marked state does not get much higher than the value of the initial coefficient. At this level, there is no way to distinguish the marked state from any others. This is similar to the search on a line here where we only see a small increase in probability at the marked state. This behaviour is caused by the connectivity of the structure itself. As there are no cycles in the Bethe lattice, the probability is split between the `branches' of the structure and so only a portion of the probability can converge on the marked state. This localization of probability in portions of the structure away from the marked state means the walker will never be able to coalesce at the marked state with any significant probability. As this maximum probability scales in a constant fashion with the number of vertices, $O(1)$, it makes no difference how long the search algorithm is run for. 

\begin{figure}[!htb]
    \centering
	\includegraphics[width=0.5\textwidth]{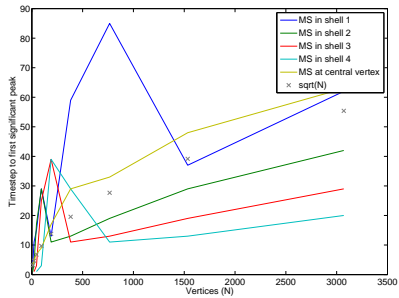}
	\caption{Time step at which the first significant peak in the probability of being at the marked state occurs for different sized data sets plotted against N, for a Bethe lattice of degree three, for varying positions of the marked state (MS). Also shown is $\sqrt{N}$ (crosses) for comparison.}
    \label{bethetime}
\end{figure}

The time to find the marked state also exhibits unusual characteristics. We see in fig.~\ref{bethetime} that as we move further from the central vertex, the timestep to the first significant peak actually reduces. However, as already mentioned, the probability at this point is so low that the marked state could never be distinguished. As the marked state only accumulates a small portion of the probability, the time to get to this amount would get faster, hence the decrease in time to `find' the marked state. We note that the `kinks' in this graph are most likely due to finite size effects. 

\begin{figure}[!htb]
    \centering
	\includegraphics[width=0.5\textwidth]{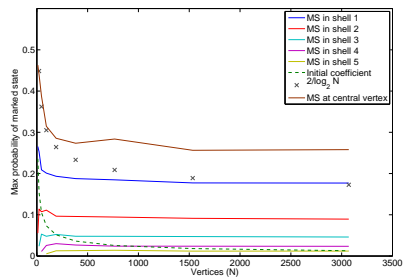}
	\caption{Maximum of the first peak in the probability of being at the marked state for different sized data sets on a Bethe lattice of degree three, plotted against N (solid), for varying positions of the marked state (MS). Also shown is the initial coefficient (dashes) and $2/\log_{2}N$ (crosses) for comparison.} 
    \label{betheprob}
\end{figure}

\section{Discussion and further work}
\label{sec:conc}

We have investigated numerically in some detail how the quantum walk search algorithm of
\citeauthor{shenvi02a} \cite{shenvi02a} behaves on several variations of the two dimensional lattice,
where it finds the marked state efficiently, and on the line and Bethe lattice, where it
does not. Our numerical results match the scaling of the algorithm on the two dimensional lattice of $O(\sqrt{N \log N})$ which, for a quantum walk approach, has been proved to be optimal \cite{magniez08a}. Although this run time matches the proven optimal scaling and the modified approach of Tulsi \cite{tulsi08a}, it is possible that our numerical results do not encompass a large enough number of vertices to see full scaling behaviour.

We studied the quantum walk search algorithm on variants of the basic 2D Cartesian lattice with varying connectivity. Specifically, these were the hexagonal lattice and the 2D Cartesian lattice with diagonal links added. This allowed the study of additional connectivity on the efficiency of the search algorithm without altering the spatial dimension. Our numerical results show the same algorithmic scaling as in the 2D Cartesian lattice case but with varying prefactors in both the time to find the marked state and also the maximum probability of the marked state. As we would expect intuitively, as we increase the connectivity of the structure, the maximum probability of the marked state also increases. Similarly, the time to find the marked state decreases with the additional connectivity. We can conclude from these results that the search algorithm is not just dependent on spatial dimension but also on the connectivity and symmetry of the data structure. Increasing the connectivity for the same spatial dimension decreases the time to find the marked state and increases the maximum probability the marked state can reach.

\begin{figure}[!tb]
    \centering
	\includegraphics[width=0.6\textwidth]{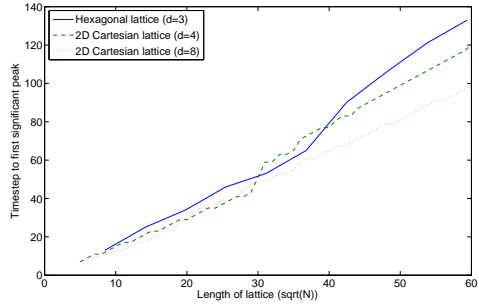}
	\caption{Time step at which the first significant peak in the probability of being at the marked state occurs for different sized data sets plotted against $\sqrt{N}$. The hexagonal lattice (solid) is shown along with the 2D Cartesian lattice (dashes) and 2D Cartesian lattice with diagonal links (dotted).}
    \label{unusual}
\end{figure}

Our numerical results also show a kink in the data at varying points depending on the structure being studied. We find this kink in the scaling of the time to run the search algorithm in all the regular, symmetric structures we have studied so far. It is most apparent in the two-dimensional Cartesian lattice case but also appears in the degree eight and hexagonal lattices. Figure \ref{unusual} shows these three cases together for comparison. We note this kink seems to occur at roughly the same number of edges, $4 \times 32^{2}$, for all tested structures. The kink seems to be due to variations in the shape of the probability of the marked state, fig.~\ref{maxprob2d}. As such, it is a numerical artifact of how we detected the maximum probability of the marked state.

Further work will include varying both the connectivity and regularity of the structures to identify in detail how these affect the efficiency of the algorithm. Disordered lattices will be studied by removing a proportion, $p$, of the edges to form a percolation lattice. We will also investigate the dependence on spatial dimension in more depth by studying fractal structures, which have non-integer values of spatial dimension, to interpolate between the one dimensional and the three dimensional cases. Percolation lattices, at their critical point, have a self similarity and so also provide an example with a fractal dimension. 

\paragraph{Acknowledgments:}
NL is funded by the UK Engineering and Physical Sciences Research Council and QNET - EPSRC network on the semantics of quantum computing. MT was funded by a Nuffield Foundation Science Undergraduate Research Bursary. ME was funded by the University of Leeds. 
VK is funded by a Royal Society University Research Fellowship.

%
%

\end{document}